\documentclass[bibyear]{aa} 

\usepackage{lscape}
\usepackage{float}
\usepackage{graphicx}

\usepackage{hyperref}
\usepackage{breakurl}
\usepackage{txfonts}
\begin{document} 

\title{APEX observations of ortho-H$_2$D$^+$ towards dense cores in the Orion B9 filament\thanks{Based on observations with the Atacama Pathfinder EXperiment (APEX) telescope under programme 0104.C-0004(A). APEX is a collaboration between the Max-Planck-Institut f{\"u}r Radioastronomie, the European Southern Observatory, and the Onsala Space Observatory.}\fnmsep\thanks{The reduced \textit{o}-H$_2$D$^+$ spectra shown in Fig.~2 are available in electronic form at the CDS via anonymous ftp to {\tt cdsarc.u-strasbg.fr} ({\tt 130.79.128.5}) or via {\tt http://cdsweb.u-strasbg.fr/cgi-bin/qcat?J/A+A/}.}}

   \author{O.~Miettinen}

   \institute{Academy of Finland, Hakaniemenranta 6, P.O. Box 131, FI-00531 Helsinki, Finland \\ \email{oskari.miettinen@aka.fi}}

   \date{Received ; accepted}

\authorrunning{Miettinen}
\titlerunning{H$_2$D$^+$ in Orion~B9}

\abstract{Initial conditions and very early stages of star formation can be probed through spectroscopic observations of deuterated molecular species.}{We aim to determine the \textit{ortho}-H$_2$D$^+$ properties (e.g. column density and fractional abundance with respect to H$_2$) in a sample of dense cores in the Orion~B9 star-forming filament, and to compare those with the previously determined source characteristics, in particular with the gas kinetic temperature, [N$_2$D$^+$]/[N$_2$H$^+$] deuterium fractionation, and level of CO depletion.}{We used the Atacama Pathfinder EXperiment (APEX) telescope to observe the 372~GHz \textit{o}-H$_2$D$^+(J_{K_a,\,K_c}=1_{1,\,0}-1_{1,\,1})$ line towards three prestellar cores and three protostellar cores in Orion~B9. We also employed our previous APEX observations of C$^{17}$O, C$^{18}$O, N$_2$H$^+$, and N$_2$D$^+$ line emission, and 870~$\mu$m dust continuum emission towards the target sources.}{The \textit{o}-H$_2$D$^+(1_{1,\,0}-1_{1,\,1})$ line was detected in all three prestellar cores, but in only one of the protostellar cores. The corresponding \textit{o}-H$_2$D$^+$ abundances were derived to be $\sim(12-30)\times10^{-11}$ and $\sim6\times10^{-11}$. Two additional spectral lines, DCO$^+(5-4)$ and N$_2$H$^+(4-3)$, were detected in the observed frequency bands with high detection rates of $100\%$ and $83\%$, respectively. We did not find any significant correlations among the explored parameters, although our results are mostly consistent with theoretical expectations. Also, the Orion~B9 cores were found to be consistent with the relationship between the \textit{o}-H$_2$D$^+$ abundance and gas temperature obeyed by other low-mass dense cores. The \textit{o}-H$_2$D$^+$ abundance was found to decrease as the core evolves.}{The \textit{o}-H$_2$D$^+$ abundances in the Orion~B9 cores are in line with those found in other low-mass dense cores and larger than derived for high-mass star-forming regions. The higher \textit{o}-H$_2$D$^+$ abundance in prestellar cores compared to that in cores hosting protostars is to be expected from chemical reactions where higher concentrations of gas-phase CO and elevated gas temperature accelerate the destruction of H$_2$D$^+$. The validity of using the [\textit{o}-H$_2$D$^+$]/[N$_2$D$^+$] abundance ratio as an evolutionary indicator, which has been proposed for massive clumps, remains to be determined when applied to these target cores. Similarly, the behaviour of the [\textit{o}-H$_2$D$^+$]/[DCO$^+$] ratio as the source evolves was found to be ambiguous. Still larger samples and observations of additional deuterated species are needed to explore these potential evolutionary indicators further. The low radial velocity of the line emission from one of the targeted prestellar cores, SMM~7 ($\sim3.6$~km~s$^{-1}$ versus the systemic Orion~B9 velocity of $\sim9$~km~s$^{-1}$), suggests that it is a chance superposition seen towards Orion~B9. Overall, as located in a dynamic environment of the Orion~B molecular cloud, the Orion~B9 filament provides an interesting target system to investigate the deuterium-based chemistry, and further observations of species like \textit{para}-H$_2$D$^+$ and D$_2$H$^+$ would be of particular interest.}


   \maketitle
%

\section{Introduction}

Deuterated molecules are very useful tracers of dense molecular cloud cores, which are direct progenitors of new stars. 
This is because at the high gas densities ($\sim10^5-10^6$~cm$^{-3}$) and low temperatures ($\sim10$~K) of the interiors of such cloud cores, many observable molecules (e.g. CO) freeze-out onto dust grains, which lowers their gas-phase abundance, rendering 
them unobservable (see e.g. \cite{bergin2007} for a review). An important exception is the trihydrogen cation, H$_3^+$, and its isotopologic forms (e.g. \cite{walmsley2004}). However, because of their symmetry, H$_3^+$ and D$_3^+$ do not posses a permanent electric dipole moment, and are therefore not observable in the conditions of dense cores. On the other hand, as asymmetric species, both H$_2$D$^+$ and D$_2$H$^+$ have dipole moments, and therefore exhibit rotational transitions that produce observable photons from dense cores. The origin of H$_2$D$^+$, and deuterium fractionation in general, lies in the exothermic reaction 

\begin{equation}
\label{eqn:deut}
{\rm H_3^+}+{\rm HD}\rightleftharpoons {\rm H_2D^+} + {\rm H_2} + \Delta E\,,
\end{equation}
where $\Delta E = 232$~K when the species lie in their lowest energy (ground) states (e.g. \cite{dalgarno1984}; \cite{gerlich2002}; \cite{sipila2017}). 

Following the first unsuccessful attempts to detect H$_2$D$^+$ in the interstellar medium (ISM; \cite{angerhofer1978}; \cite{phillips1985}; \cite{pagani1992}; \cite{vandishoeck992}), the astronomical detectability of H$_2$D$^+$ was finally demonstrated for the first time about 20~yr ago by Stark et al. (1999) who detected the \textit{ortho}-H$_2$D$^+(J_{K_a,\,K_c}=1_{1,\,0}-1_{1,\,1})$ line at 372~GHz towards the low-mass star-forming core NGC~1333 IRAS~4A. So far, \textit{o}-H$_2$D$^+$ has only been detected towards approximately 15 different low-mass dense cores (e.g. \cite{harju2008}; \cite{caselli2008}; \cite{friesen2010}, 2014) and a few high-mass star-forming regions (\cite{swift2009}; \cite{pillai2012}; \cite{giannetti2019}). Reaching a better understanding of H$_2$D$^+$-based deuterium chemistry in dense cores at different evolutionary stages (e.g. starless cores vs. protostellar cores) requires a more homogeneous H$_2$D$^+$ survey of dense cores in a common parent molecular cloud than those published so far. With such data, the potential role played by the different initial chemical conditions and distance (that is angular resolution) effects could be (largely) eliminated, which in turn is expected to yield more comparable results between dense cores of different ages. Motivated by this possibility, we used the Atacama Pathfinder EXperiment (APEX\footnote{\url{http://www.apex-telescope.org/}}; \cite{gusten2006}) telescope to carry out an \textit{o}-H$_2$D$^+(1_{1,\,0}-1_{1,\,1})$ survey towards six dense cores (three prestellar cores and three protostellar cores) in the Orion~B9 star-forming region.

Miettinen et al. (2009) mapped the Orion~B9 region at $\lambda=870$~$\mu$m using the Large APEX BOlometer CAmera (LABOCA; \cite{siringo2009}) on APEX. These latter authors identified 12 dense cores in total, of which 6 were found to be associated with \textit{Spitzer} 24~$\mu$m sources, and were therefore deemed protostellar. The remaining cores have no mid-infrared counterparts and are likely starless. We note that of the present target sources, four (SMM~1, 3, 6, and 7) were first uncovered by our LABOCA mapping (of which only SMM~3 has a 24~$\mu$m counterpart). The remaining two sources included in the present study, IRAS~05399-0121 and IRAS~05405-0117 (hereafter, IRAS~05399 and IRAS~05405), are associated with infrared point sources that were previously identified with the \textit{Infrared Astronomical Satellite} (\textit{IRAS}; \cite{neugebauer1984}). 

The physical and chemical properties of the Orion~B9 cores (e.g. the NH$_3$-based gas kinetic temperature ($T_{\rm kin}$), CO depletion factor ($f_{\rm D}({\rm CO})$), and the level of N$_2$H$^+$ deuteration, i.e. the $[{\rm N_2D^+}]/[{\rm N_2H^+}]$ ratio) were further characterised by Miettinen et al. (2010, 2012) through Effelsberg and APEX molecular line observations. Miettinen et al. (2010) used a virial parameter analysis to examine whether the starless cores in the region are gravitationally bound ($\alpha_{\rm vir} <2$), and therefore prestellar in nature. On the basis of this, all the starless cores in the present sample can be considered prestellar.

Miettinen et al. (2012) imaged the Orion~B9 region at $\lambda=350$~$\mu$m using the Submillimetre APEX BOlometer CAmera (SABOCA; \cite{siringo2010}). The 350~$\mu$m imaging revealed that some of the Orion~B9 cores (e.g. SMM~3 and SMM~6) host two or more subfragments (see also \cite{miettinen2013a},b). As part of the Orion~B (L1630) molecular cloud, Orion~B9 was mapped with \textit{Herschel} (\cite{pilbratt2010}) in the \textit{Herschel} Gould Belt Survey (HGBS; \cite{andre2010})\footnote{The HGBS is a \textit{Herschel} key programme jointly carried out by SPIRE Specialist Astronomy Group 3 (SAG 3), scientists of several institutes in the PACS Consortium (CEA Saclay, INAF-IFSI Rome and INAF-Arcetri, KU Leuven, MPIA Heidelberg), and scientists of the \textit{Herschel} Science Center (HSC). For more details, see \url{http://gouldbelt-herschel.cea.fr}.}. The \textit{Herschel} images revealed that Orion~B9 is actually a filamentary system in which the present target cores appear to be embedded. Miettinen (2012) found that there is a sharp velocity gradient in the parent filament (across its short axis), and suggested that it might be a manifestation of a shock front, resulting from the feedback from the expanding \ion{H}{II} region and OB cluster NGC~2024 that lies
about 4~pc to the southwest of Orion~B9. Ohama et al. (2017), who mapped the NGC~2024 region in $^{13}$CO$(J=2-1)$, found that NGC~2024 is comprised of two velocity components (at 9.5~km~s$^{-1}$ and 11.5~km~s$^{-1}$). The authors suggested that collision between two clouds has triggered the formation of stars in NGC~2024, with a collision timescale of $2 \times 10^5$~yr. The systemic radial velocity of the lower velocity component (9.5~km~s$^{-1}$) is consistent with that of Orion~B9 ($\sim9$~km~s$^{-1}$), which suggests that the clouds lie at the same distance from the Sun. It is therefore possible that the formation of dense cores and stars in
Orion~B9 (or even that of the whole filament) was triggered by external positive feedback from NGC~2024. However, further 
studies are required to test this hypothesis. Nevertheless, Orion~B, which is located within the Barnard's Loop, is known to be affected by radiation and winds from massive OB stars (e.g. \cite{cowie1979}; \cite{bally2008} for a review; \cite{schneider2013}). This further strengthens the hypothesis that the Orion~B9 filament is indeed located in a dynamic environment where triggered star formation is plausible.

We note that Harju et al. (2006) already used APEX to search for \textit{o}-H$_2$D$^+(1_{1,\,0}-1_{1,\,1})$ emission towards three selected positions near and around IRAS~05405 (the N$_2$H$^+(1-0)$ peaks found by Caselli \& Myers (1994)), but the target positions were offset from the submillimetre dust continuum peaks that we later uncovered, rendering the line detections fairly weak ($\lesssim 3.7\sigma$). Caselli et al. (2008) also searched for \textit{o}-H$_2$D$^+(1_{1,\,0}-1_{1,\,1})$ emission towards a position near ($10\farcs4$ offset) to one of those observed by Harju et al. (2006), but no line was detected. 

In this paper, we present the results of our 372~GHz \textit{o}-H$_2$D$^+$ observations towards dense cores in Orion~B9. The observations and data reduction are described in Sect.~2. The analysis and its results are presented in Sect.~3. In Sect.~4, we discuss the results, and in Sect.~5 we summarise our results and present our main conclusions. Throughout this paper we adopt a distance of $d=420$~pc to Orion~B9, which is consistent with the most recent distance measurement for NGC~2024 (\cite{kounkel2017}).

\section{Source sample, observations, and data reduction}

\begin{figure*}[!htb]
\begin{center}
\includegraphics[scale=0.75]{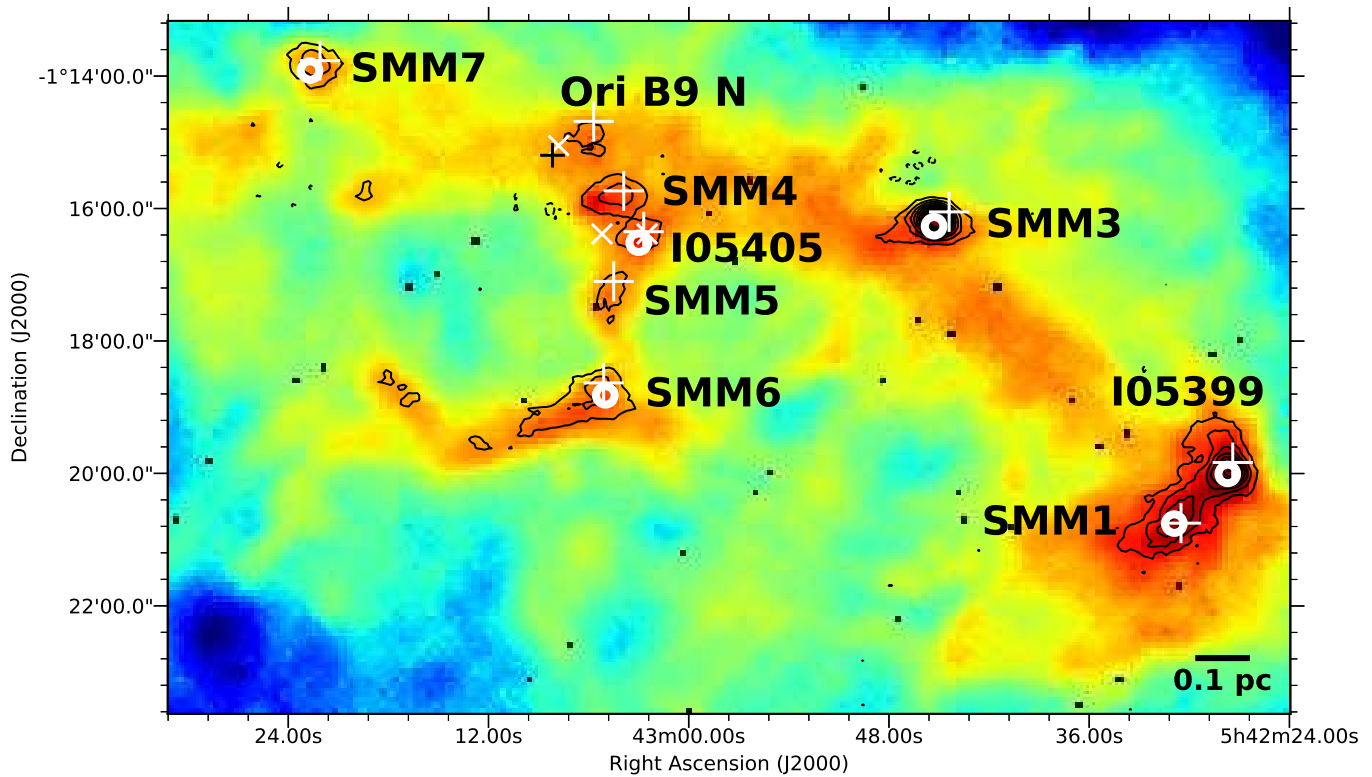}
\vspace*{-10mm}
\caption{\textit{Herschel}/SPIRE (Spectral and Photometric Imaging REceiver) 250~$\mu$m image towards Orion B9. The image is displayed using a non-linear (arcsinh) stretch, which allows us to better see the faint, extended dust emission. The overlaid contours represent the LABOCA 870~$\mu$m emission (\cite{miettinen2009}); the contours start at $3\sigma$, and increase in steps of $3\sigma$, where $3\sigma=90$~mJy~beam$^{-1}$. The circles indicate the target positions of the present \textit{o}-H$_2$D$^+$ observations 
(i.e. SABOCA 350~$\mu$m emission peaks of the cores (\cite{miettinenetal2012}; \cite{miettinen2013a})), and the circle size corresponds 
to the beam size of these observations ($16\farcs8$ HPBW). The white plus signs show the target positions of the spectral line observations from Miettinen et al. (2010, 2012), while the white crosses show the \textit{o}-H$_2$D$^+$ target positions observed 
by Harju et al. (2006). The black plus symbol marks the position from which Caselli et al. (2008) searched for \textit{o}-H$_2$D$^+$ emission. A scale bar of 0.1~pc projected length is shown in the bottom right corner.}
\label{figure:map}
\end{center}
\end{figure*}

The target sample of the present study consists of the Orion~B9 prestellar cores SMM~1, 6, and 7, and
the protostellar cores IRAS~05399, IRAS~05405, and SMM~3. Besides spanning different 
early evolutionary stages, these six sources were selected for this study because their spatial distribution
covers different parts of the Orion~B9 region (see Fig.~\ref{figure:map}). The source list is given in Table~1.

The SABOCA 350~$\mu$m peak positions of the aforementioned target sources were  
observed with the 12 m APEX telescope at the frequency of the \textit{o}-H$_2$D$^+(1_{1,\,0}-1_{1,\,1})$ transition, that is 
$\nu=372.42135580$~GHz (Cologne Database for Molecular Spectroscopy (CDMS\footnote{\url{http://www.astro.uni-koeln.de/cdms}}; \cite{muller2005})). The observations were carried out on 27-30 August and 15-16 December 2019, and the amount of precipitable water vapour
(PWV) during the observations was measured to be between 0.3~mm and 1~mm, which corresponds to a zenith atmospheric transmission 
range of about 75--43\% at the observed frequency\footnote{The APEX atmospheric transmission calculator is available at \url{http://www.apex-telescope.org/sites/chajnantor/atmosphere/transpwv/}.}.

As a front end we used the Large APEX sub-Millimetre Array (LAsMA), which is 
a seven-pixel sideband separating (2SB) heterodyne array receiver that can be tuned 
within a frequency range of 273--374~GHz (\cite{gusten2008}). The backend was a Fast Fourier Transform fourth-Generation spectrometer (FFTS4G) that consists of two sidebands. Each sideband has two spectral windows of 4~GHz bandwidth, which provides both orthogonal polarisations and leads to a total bandwidth of 8~GHz. The \textit{o}-H$_2$D$^+$ line 
was tuned in the upper sideband (USB). The 65\,536 channels of the spectrometer yielded a 
spectral resolution of 61~kHz, which corresponds to 49~m~s$^{-1}$ at the observed frequency. The beam size (half-power beam width or HPBW) at the observed frequency is $16\farcs8$, which corresponds to a physical resolution of 0.034~pc. 

The observations were performed in the wobbler-switching mode with a $50\arcsec$ azimuthal throw between two positions on sky (symmetric offsets), and a chopping rate of $R = 0.5$~Hz. The total on-source integration time (excluding overheads) was about 18--46~min depending on the source. The telescope focus and pointing were optimised and checked at regular intervals on the planet Uranus, 
the variable star \textit{o}-Ceti, the asymptotic giant branch star R Doradus, and the carbon stars W Orionis and R Leporis.  
The pointing was found to be accurate to $\sim3\arcsec$. The system temperatures during the observations were in
the range $T_{\rm sys}\simeq 326-465$~K. Calibration was made by means of the chopper-wheel technique, and the output intensity scale given by the system is the antenna temperature corrected for the atmospheric attenuation ($T_{\rm A}^{\star}$). The observed intensities were converted to the main-beam brightness temperature scale by $T_{\rm MB}=T_{\rm A}^{\star}/\eta_{\rm MB}$, 
where $\eta_{\rm MB}$ is the main-beam efficiency. The value of $\eta_{\rm MB}$ at the observed frequency was calculated as follows. The aperture efficiency at the central frequency of LAsMA ($\eta_{\rm a}=0.59$ when estimated from observations towards Jupiter) was first scaled to that at the observed frequency using the Ruze formula (\cite{ruze1952}), and then converted to 
the value of $\eta_{\rm MB}$ using the approximation formula $\eta_{\rm MB}\simeq1.2182 \times \eta_{\rm a}$. This yielded 
the values $\eta_{\rm a}=0.58$ and $\eta_{\rm MB}=0.71$.\footnote{The APEX antenna efficiences can be found at \url{ http://www.apex-telescope.org/telescope/efficiency/index.php}.} The absolute calibration uncertainty was estimated to
be about 10\% (e.g. \cite{dumke2010}).

The spectra were reduced using the Continuum and Line
Analysis Single-dish Software ({\tt CLASS90}) program of
the GILDAS software package\footnote{Grenoble Image and Line Data Analysis Software (GILDAS) is provided and actively developed by Institut de Radioastronomie Millim\'etrique (IRAM), and is available at \url{http://www.iram.fr/IRAMFR/GILDAS}.} (version mar19b). The individual spectra were averaged with weights proportional to the integration time, divided by the square of the system temperature ($w_i \propto t_{\rm int}/T_{\rm sys}^2$). The resulting spectra were smoothed using the Hann window function to a velocity resolution of 98~m~s$^{-1}$ to improve the signal-to-noise ratio. Linear (first-order) baselines were determined from the velocity ranges free of spectral line features, and then subtracted from the spectra. The resulting $1\sigma$ rms noise levels at the 98~m~s$^{-1}$ velocity resolution were 28--33~mK on a $T_{\rm MB}$ scale. The observational parameters are provided in Table~\ref{table:observations}.

In Table~\ref{table:lines}, we summarise all the spectral line observations that were employed in the present study. In addition to the newly detected spectral lines, we also used spectral line data from Miettinen et al. (2010, 2012), Miettinen \& Offner (2013b), and Miettinen (2016).

\begin{table}[H]
\renewcommand{\footnoterule}{}
\caption{Source sample.}
{\small
\begin{minipage}{1\columnwidth}
\centering
\label{table:sources}
\begin{tabular}{c c c c c}
\hline\hline 
Source & $\alpha_{2000.0}$ & $\delta_{2000.0}$ & $\Delta_{\rm line}$ & Type\\
       & [h:m:s] & [$\degr$:$\arcmin$:$\arcsec$] & [$\arcsec$] & \\ 
\hline 
IRAS 05399-0121 & 05 42 27.7 & -01 20 00.4 & 11.3 & Class 0/I \\
SMM 1 & 05 42 30.9 & -01 20 45.4 & 6.0 & prestellar \\
SMM 3 & 05 42 45.3 & -01 16 16.0 & 18.7 & Class 0 \\
IRAS 05405-0117 & 05 43 03.0 & -01 16 31.0 & 11.0 & Class 0 \\
SMM 6 & 05 43 05.0 & -01 18 49.3 & 11.4 & prestellar \\
SMM 7 & 05 43 22.7 & -01 13 55.0 & 12.7 & prestellar \\
\hline
\end{tabular} 
\tablefoot{The coordinates refer to the SABOCA 350~$\mu$m peak positions of the sources (\cite{miettinenetal2012}; \cite{miettinen2013a},b). The angular offset, $\Delta_{\rm line}$, refers to the distance from these SABOCA peaks to the target positions 
observed by Miettinen et. al (2010, 2012). The references to the evolutionary stages of the sources are Miettinen et al. (2009, 2010, 2012), Miettinen \& Offner (2013a), and Miettinen (2016).}
\end{minipage} }
\end{table}

\begin{table}[H]
\renewcommand{\footnoterule}{}
\caption{Observational parameters.}
{\normalsize
\begin{minipage}{1\columnwidth}
\centering
\label{table:observations}
\begin{tabular}{c c c c c}
\hline\hline 
Source & $t_{\rm ON}$ & PWV & $T_{\rm sys}$\tablefootmark{a} & $1\sigma$ rms\tablefootmark{a} \\
       & [min] & [mm] & [K] & [mK] \\ 
\hline 
IRAS 05399-0121 & 46.4 & 0.6-1.0 & 655 & 28 \\
SMM 1 & 32.1 & 0.6-1.0 & 600 & 33 \\
SMM 3 & 22.4 & 0.3-1.0 & 581 & 33 \\
IRAS 05405-0117 & 38.4 & 0.4-1.0 & 605 & 28\\
SMM 6 & 17.6 & 0.4-1.0 & 459 & 29 \\
SMM 7 & 20.8 & 0.4-1.0 & 468 & 29 \\
\hline
\end{tabular} 
\tablefoot{\tablefoottext{a}{The system temperature and $1\sigma$ rms noise level are given in the main-beam brightness temperature scale.}}
\end{minipage} }
\end{table}

\begin{table*}
\caption{Spectral line transitions employed in the present study.}
\begin{minipage}{2\columnwidth}
\centering
\renewcommand{\footnoterule}{}
\label{table:lines}
\begin{tabular}{c c c c c c c}
\hline\hline
Transition & $\nu$ & \multicolumn{2}{c}{HPBW} & \multicolumn{2}{c}{$\Delta_{\rm o-H_2D^+}$\tablefootmark{a}} & Reference \\
           & [MHz] & [$\arcsec$] & [pc] & [$\arcsec$] & [pc] & \\
\hline 
\textit{o}-H$_2$D$^+(1_{1,\,0}-1_{1,\,1})$ & 372\,421.35580 & 16.8 & 0.034 & 0 & 0 & This work\\
N$_2$H$^+(4-3)$ & 372\,672.526 & 16.7 & 0.034 & 0 & 0 & This work\\
DCO$^+(5-4)$ & 360\,169.7771 & 17.3 & 0.035 & 0 & 0 & This work\\
C$^{18}$O$(2-1)$ & 219\,560.3568 & 28.4 & 0.058 & 2.2 & 0.004 & \cite{miettinen2016}\tablefootmark{b}\\
C$^{17}$O$(2-1)$ & 224\,714.199 & 27.8 & 0.057 & 0 & 0 & \cite{miettinen2013b}\tablefootmark{c}\\
                 &    \ldots          &  \ldots     &   \ldots     & 6.0-18.7 & 0.012-0.038 & \cite{miettinenetal2012}\\
N$_2$D$^+(3-2)$ & 231\,321.912 & 27.0 & 0.055 & 0 & 0 & \cite{miettinen2013b}\tablefootmark{c}\\
				&     \ldots          &   \ldots    &   \ldots     & 6.0-18.7 & 0.012-0.038 & \cite{miettinenetal2012}\\
\textit{p}-NH$_3(1,\,1)$ & 23\,694.4955 & 40 & 0.081 & 6.0-18.7 & 0.012-0.038 & \cite{miettinen2010}\\
\textit{p}-NH$_3(2,\,2)$ & 23\,722.6333 & 40 & 0.081 & 6.0-18.7 & 0.012-0.038 & \cite{miettinen2010}\\
\hline
\end{tabular} 
\tablefoot{\tablefoottext{a}{Offset with respect to the present \textit{o}-H$_2$D$^+$ target positions. When this offset is different for different target source, a range of values is quoted (cf.~Table~\ref{table:sources}).}\tablefoottext{b}{The C$^{18}$O$(2-1)$ line was observed only towards SMM~3.}\tablefoottext{c}{Miettinen \& Offner (2013b) studied the prestellar core SMM~6 (their source SMM~6a corresponds to the present target position).}}
\end{minipage} 
\end{table*}

\section{Analysis and results}

\subsection{Spectral line analysis}

\begin{figure*}[!htb]
\begin{center}
\includegraphics[width=0.33\textwidth]{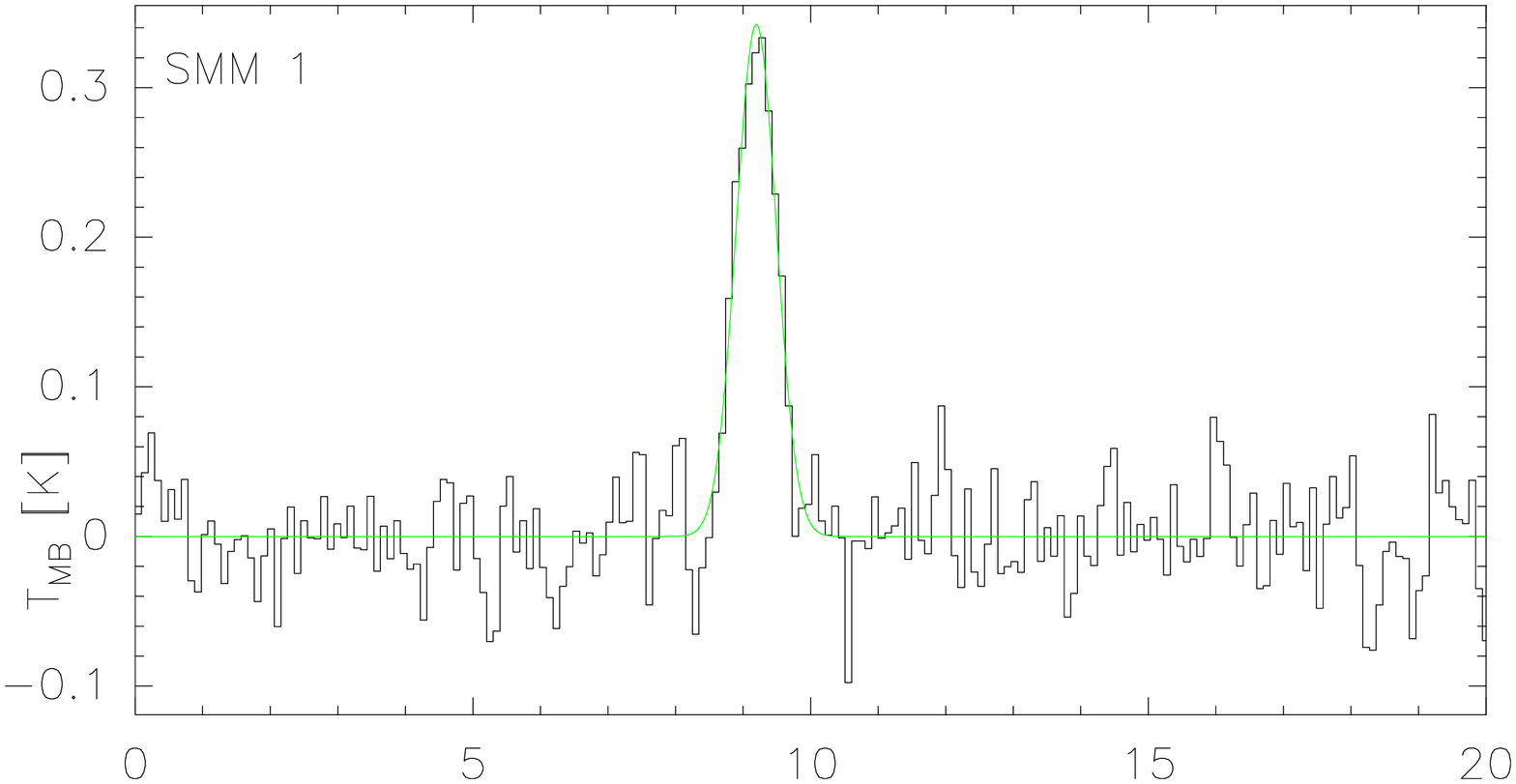}
\includegraphics[width=0.33\textwidth]{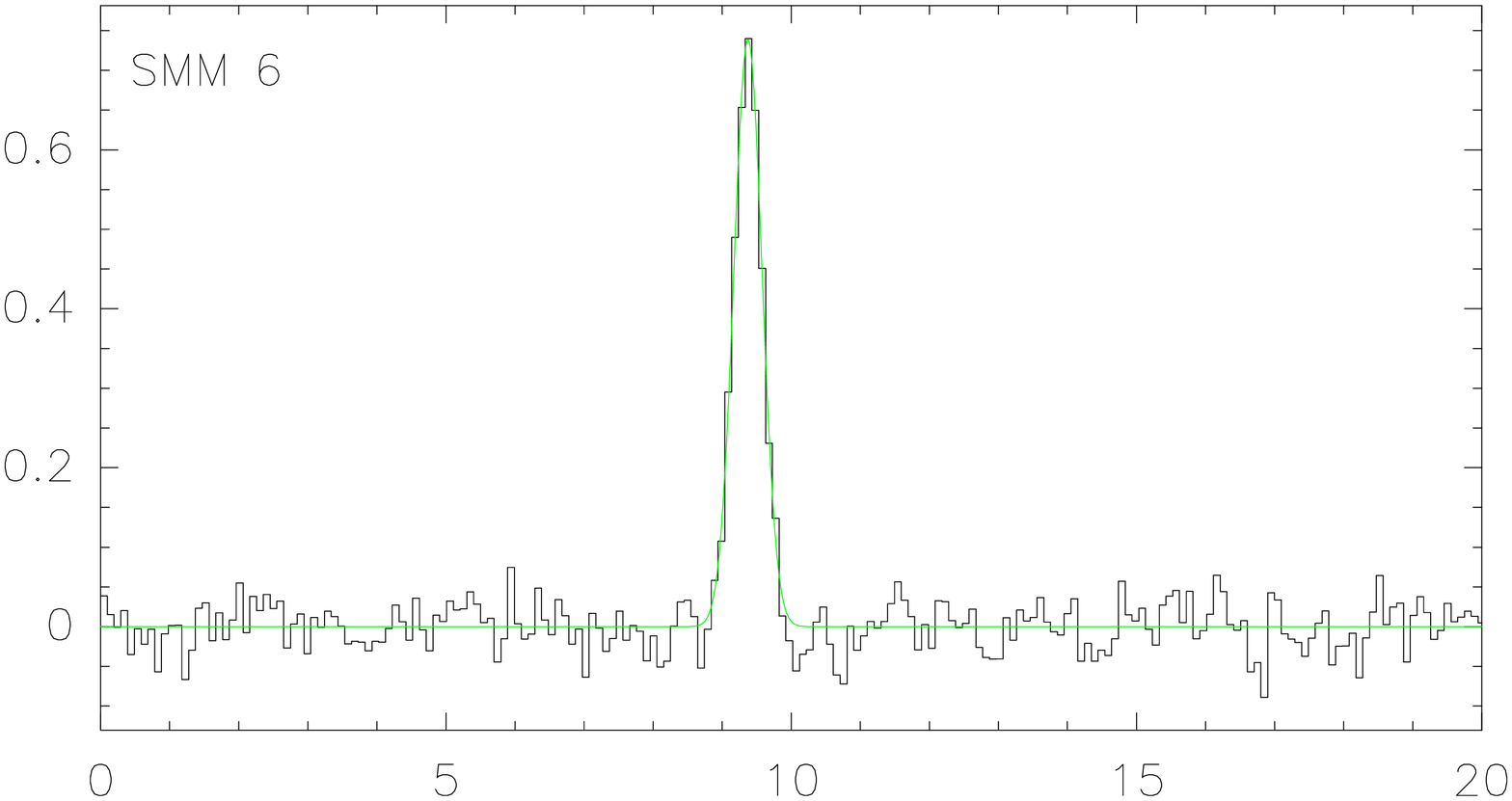}
\includegraphics[width=0.33\textwidth]{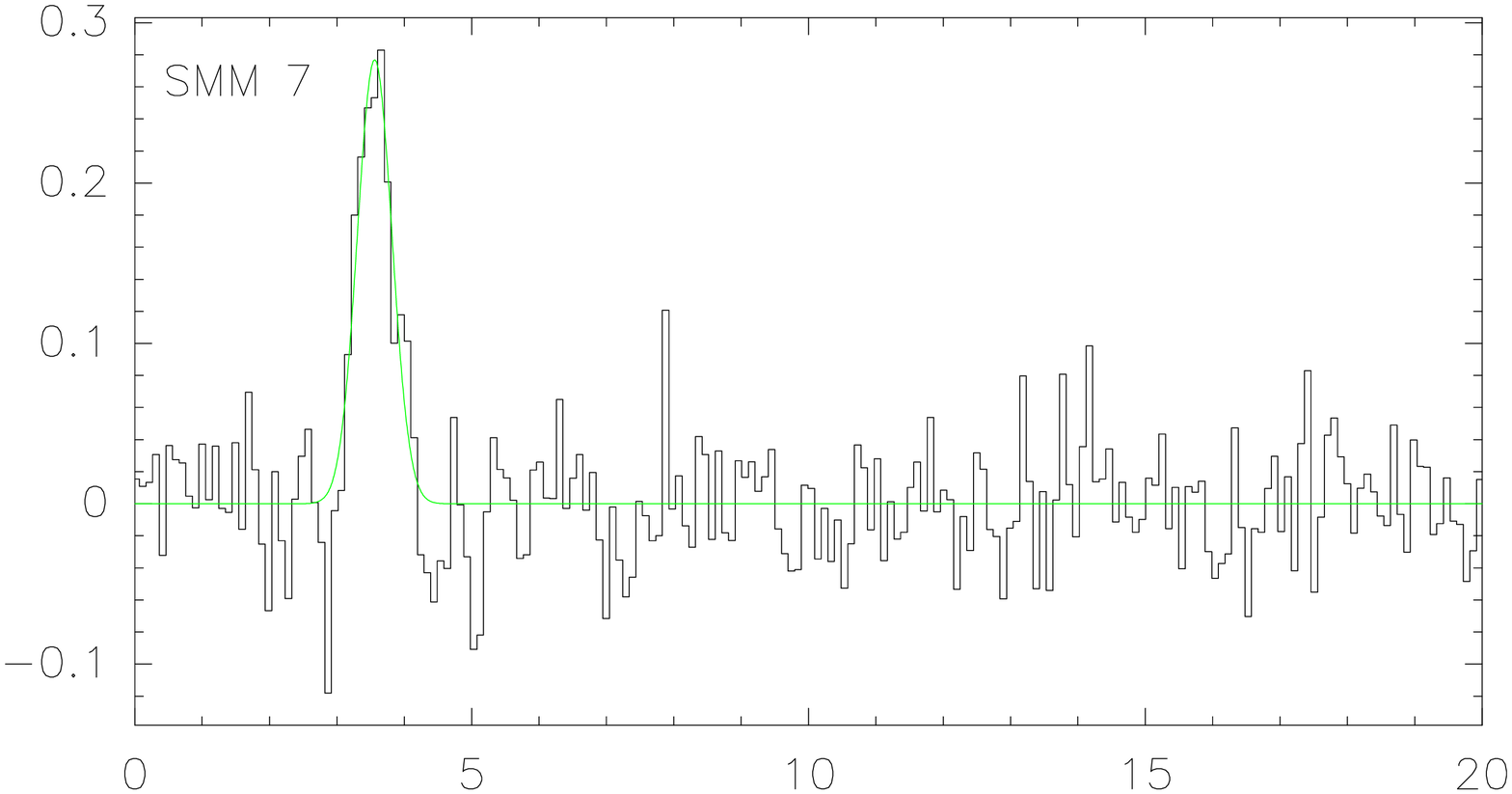}
\includegraphics[width=0.33\textwidth]{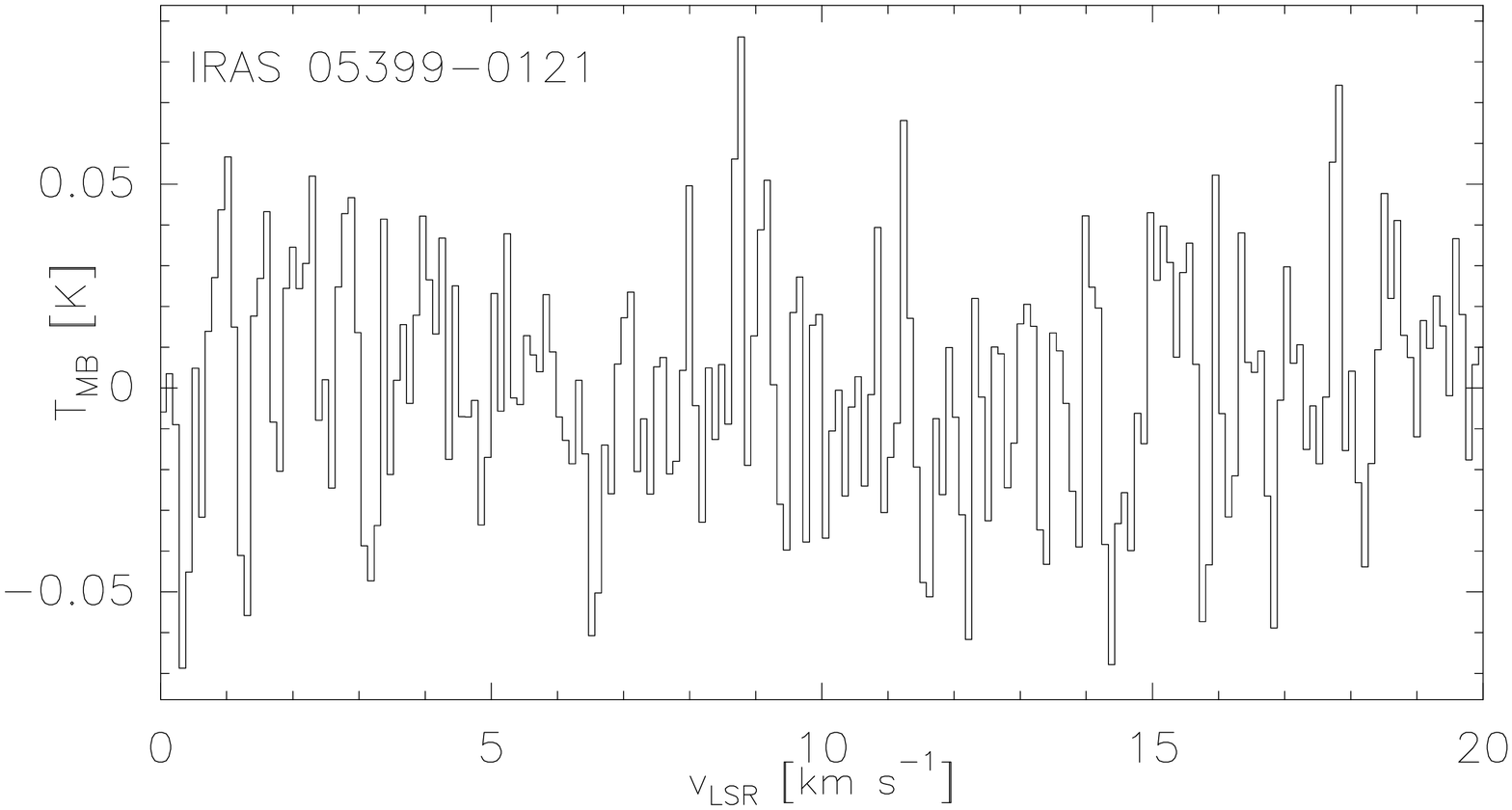}
\includegraphics[width=0.33\textwidth]{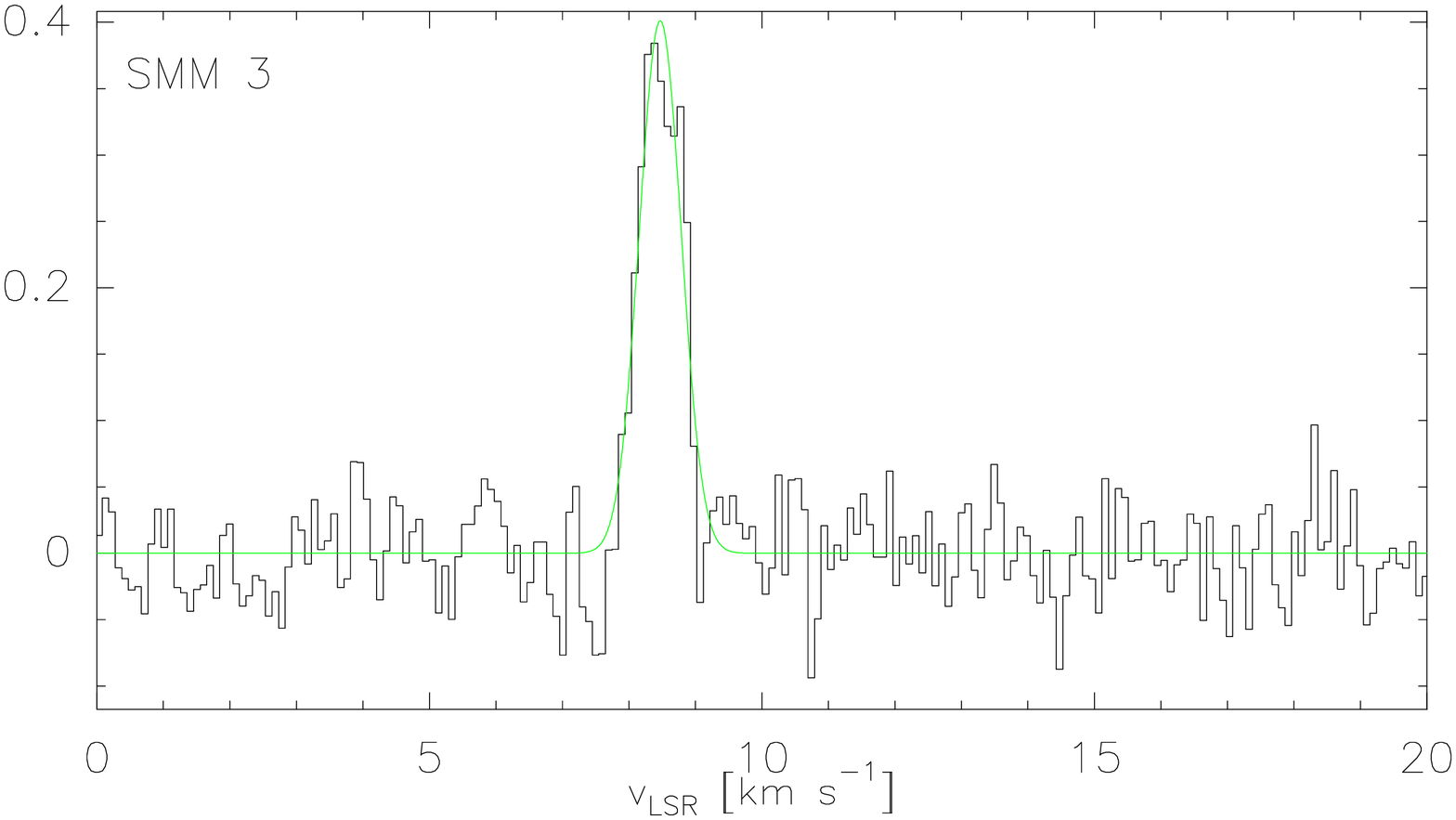}
\includegraphics[width=0.33\textwidth]{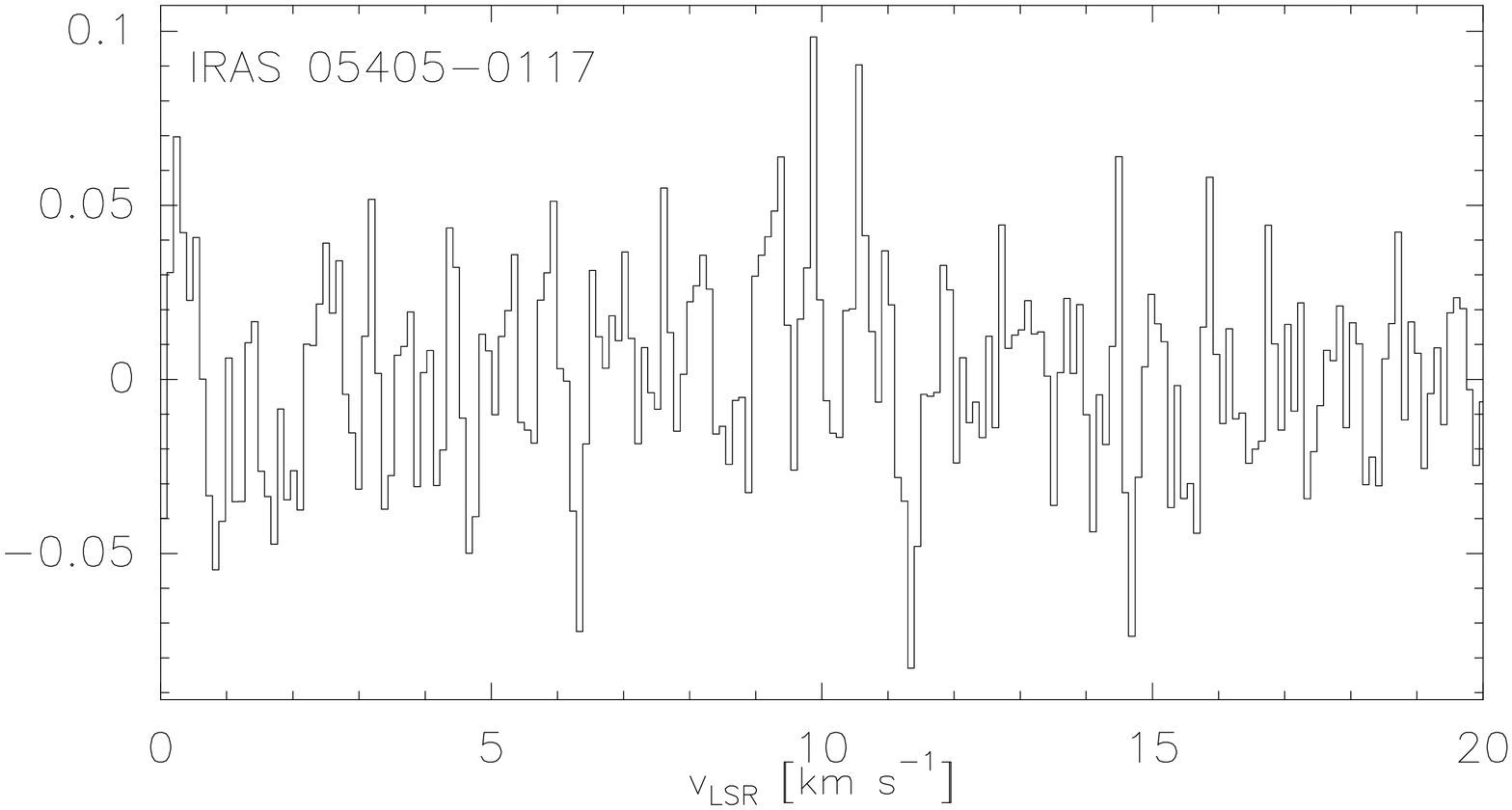}
\caption{APEX \textit{o}-H$_2$D$^+(1_{1,\,0}-1_{1,\,1})$ spectra towards the prestellar cores (\textit{top row}) and protostellar cores (\textit{bottom row}) in our sample. Gaussian fits to the lines are overlaid in green. While the velocity range shown in each panel is the same, the intensity range is different to better show the line profiles.}
\label{figure:spectra}
\end{center}
\end{figure*}

The \textit{o}-H$_2$D$^+(1_{1,\,0}-1_{1,\,1})$ spectra are shown in Fig.~\ref{figure:spectra}. The detected lines 
were fitted by single Gaussian functions using {\tt CLASS90}. The obtained line parameters are listed in Table~\ref{table:parameters}. In addition to the formal $1\sigma$ fitting errors output by {\tt CLASS90}, the errors in the peak intensity ($T_{\rm MB}$) and the integrated intensity of the line ($\int T_{\rm MB} {\rm d}v$) also include the 10\% calibration uncertainty, which were added in quadrature.

The observed frequency bands revealed the presence of two additional spectral lines towards the target sources. The N$_2$H$^+(4-3)$ line at 372\,672.526~GHz in the USB was detected towards four target sources, while DCO$^+(5-4)$ at 360\,169.7771 GHz in the lower sideband (LSB) was detected in all of our target sources (see the second and third row in Table~\ref{table:lines}). These additional line detections and their analysis are presented in more detail in Appendix~A.

\begin{table*}
\caption{\textit{o}-H$_2$D$^+(1_{1,\,0}-1_{1,\,1})$ spectral line parameters.}
\begin{minipage}{2\columnwidth}
\centering
\renewcommand{\footnoterule}{}
\label{table:parameters}
\begin{tabular}{c c c c c c c}
\hline\hline 
Source & $v_{\rm LSR}$ & $\Delta v_{\rm LSR}$ & $T_{\rm MB}$ & $\int T_{\rm MB} {\rm d}v$ & $\tau$\tablefootmark{a} & $T_{\rm ex}$\tablefootmark{b}\\
       & [km s$^{-1}$] & [km s$^{-1}$] & [K] & [K km s$^{-1}$] & & [K]  \\
\hline
IRAS 05399-0121 & \ldots & \ldots & $<0.09$\tablefootmark{c} & $<0.07$\tablefootmark{d} & $<0.02$ & $13.5\pm1.6$\\ [1ex] 
SMM 1 & $9.20\pm0.02$ & $0.70\pm0.04$ & $0.34\pm0.05$ & $0.25\pm0.03$ & $0.07\pm0.01$ & $11.9\pm0.9$ \\ [1ex] 
SMM 3 & $8.47\pm0.02$ & $0.74\pm0.03$ & $0.40\pm0.05$ & $0.32\pm0.03$ & $0.09\pm0.01$ & $11.3\pm0.8$ \\ [1ex] 
IRAS 05405-0117 & \ldots & \ldots & $<0.09$\tablefootmark{c} & $<0.07$\tablefootmark{d} & $<0.02$ & $11.3\pm0.6$ \\ [1ex] 
SMM 6 & $9.38\pm0.01$ & $0.48\pm0.01$ & $0.74\pm0.08$ & $0.38\pm0.04$ & $0.19\pm0.02$ & $11.0\pm0.4$ \\ [1ex]      
SMM 7 & $3.56\pm0.02$ & $0.60\pm0.04$ & $0.28\pm0.04$ & $0.18\pm0.02$ & $0.09\pm0.03$ & $9.4\pm1.1$\\ [1ex] 
\hline 
\end{tabular} 
\tablefoot{Columns (2)--(5) give the local standard of rest (LSR) radial velocity ($v_{\rm LSR}$), full width at half maximum (FWHM; $\Delta v$), peak intensity ($T_{\rm MB}$), and the integrated intensity of the line ($\int T_{\rm MB} {\rm d}v$).\tablefoottext{a}{Peak optical thickness calculated using $T_{\rm ex}$ from the last column.}\tablefoottext{b}{The value of $T_{\rm ex}$ was assumed to be equal to $T_{\rm kin}({\rm NH_3})$ (\cite{miettinen2010}).}\tablefoottext{c}{A $3\sigma$ upper limit is placed for the non-detections.}\tablefoottext{d}{The integrated intensity upper limit for these protostellar cores was estimated as the intensity upper limit ($<0.09$~K) times the detected linewidth in the Class~0 object SMM~3 (0.74~km~s$^{-1}$).}         }
\end{minipage}
\end{table*}

\subsection{Molecular column densities and fractional abundances}

The beam-averaged \textit{o}-H$_2$D$^+$ column densities were calculated by using
the standard, local thermodynamic equilibrium (LTE)-based formula, 

\begin{equation}
\label{eqn:CD}
N=\frac{3h \epsilon_0}{2\pi^2}\frac{1}{\mu^2S}\frac{Z_{\rm rot}(T_{\rm ex})}{g_Kg_I}e^{E_{\rm u}/k_{\rm B}T_{\rm ex}}F(T_{\rm ex})\int \tau(v){\rm d}v\, ,
\end{equation}
where $h$ is the Planck constant, $\epsilon_0$ is the vacuum permittivity, $\mu$ is the permanent electric dipole moment, $S$ is the line strength, $Z_{\rm rot}$ is the partition function, $T_{\rm ex}$ is the excitation temperature, 
$g_K$ is the $K$-level degeneracy, $g_I$ is the reduced nuclear spin degeneracy, $E_{\rm u}$ is the upper-state energy, 
$k_{\rm B}$ is the Boltzmann constant, the function $F(T_{\rm ex})$ is defined as $F(T_{\rm ex})=\left(e^{h\nu /k_{\rm B}T_{\rm ex}}-1\right)^{-1}$, and, for a Gaussian profile, the last integral term can be expressed as 

\begin{equation}
\int \tau(v){\rm d}v = \frac{\sqrt{\pi}}{2\sqrt{\ln 2}}\Delta v \tau_0 \simeq 1.064\Delta v \tau_0\, ,
\end{equation}
where $\tau_0$ is the peak optical thickness of the line. 

The value of the product $\mu^2S$ for the observed \textit{o}-H$_2$D$^+$ transition, $1.62006$~D$^2$, 
was taken from the Splatalogue database\footnote{\url{http://www.cv.nrao.edu/php/splat/}} (the CDMS value therein). 
We assumed that the \textit{o}-H$_2$D$^+(1_{1,\,0}-1_{1,\,1})$ transition is thermalised at the NH$_3$-based gas kinetic 
temperature derived by Miettinen et al. (2010) towards the target sources, that is 
$T_{\rm ex}(o-{\rm H_2D^+})=T_{\rm kin}({\rm NH_3})$; see Table~\ref{table:parameters}.  
This assumption is supported by the results of Caselli et al. (2008), where the $T_{\rm kin}/T_{\rm ex}(o-{\rm H_2D^+})$ ratio 
for their sample ranges from 1 to 1.47 (at a critical density of $10^5$~cm$^{-3}$), with a mean (median) of 1.1 (1.0). 
The partition function values at the aforementioned $T_{\rm ex}$ values were interpolated from 
the values computed by Friesen et al. (2010; Table~3 therein). Because H$_2$D$^+$ is an asymmetric top molecule (the 
rotational constants from the CDMS database yield a Ray's asymmetry parameter of $\kappa=-0.0657$ (\cite{ray1932})), 
the value of $g_K$ is unity (i.e. no $K$-level degeneracy), while $g_I=3/4$ for the observed \textit{ortho} form of H$_2$D$^+$ (\cite{turner1991}). The value of $E_{\rm u}/k_{\rm B}$ is 17.8~K (e.g. \cite{friesen2010}). The values of $\tau_0$ 
were calculated from the antenna equation ($T_{\rm MB}\propto (1-e^{-\tau})$; see e.g. Eq.~(1) in \cite{miettinen2009})
using the aforementioned NH$_3$-based $T_{\rm ex}$ values and assuming that the
background temperature is equal to that of the cosmic microwave background (CMB) radiation, that is $T_{\rm bg}=T_{\rm CMB}=2.725$~K 
(\cite{fixsen2009}). 

The fractional abundances of \textit{o}-H$_2$D$^+$ were calculated by dividing the \textit{o}-H$_2$D$^+$ column density 
by the H$_2$ column density, that is $x = N/N({\rm H_2})$. The $N({\rm H_2})$ values were derived
from our LABOCA dust continuum data re-reduced by Miettinen (2016). The angular resolution of the corresponding LABOCA image, 
$19\farcs86$ (HPBW), is comparable to the beam size of our \textit{o}-H$_2$D$^+$ observations (only a factor of 1.18 difference).  
We employed the standard formula that relates $N({\rm H_2})$ with the dust peak surface brightness (see e.g. Eq.~(3) in \cite{miettinen2009}). We assumed that the dust temperature is equal to $T_{\rm kin}({\rm NH_3})$ derived by Miettinen et al. (2010), the mean molecular weight per H$_2$ molecule is 2.82, the dust opacity at 870~$\mu$m is 1.38~cm$^2$~g$^{-1}$, and that 
the dust-to-gas mass ratio is $R_{\rm dg}=1/141$ (see \cite{miettinen2016} and references therein for the details). 

In the present study, we also employ the C$^{17}$O, N$_2$H$^+$, and N$_2$D$^+$ data from Miettinen et al. (2012) and Miettinen \& Offner (2013b), and C$^{18}$O data from Miettinen (2016; only for SMM~3). The column densities of these linear molecules were also re-calculated using Eq.~(\ref{eqn:CD}) so that they can be compared with the present \textit{o}-H$_2$D$^+$ results (e.g. Miettinen et al. (2012) employed a non-LTE approach in their column density analysis). The C$^{17}$O and C$^{18}$O lines were assumed to be thermalised at $T_{\rm kin}({\rm NH_3})$, while the $T_{\rm ex}[{\rm NH_3(1,\, 1)}]$ values were used for N$_2$H$^+$ and N$_2$D$^+$ (we refer to \cite{miettinen2016} for a more detailed description). To properly calculate the fractional abundances of the aforementioned linear molecules, the corresponding $N({\rm H_2})$ values were derived from the LABOCA data smoothed to the coarser resolution of the line observations ($22\farcs3-28\farcs4$ HPBW). The beam-averaged column densities and abundances with respect to H$_2$ calculated in this section are given in Table~\ref{table:results}. The column densities and fractional abundances calculated from the additional spectral line detections are presented in Appendix~A.

\begin{table*}
\caption{Molecular column densities and fractional abundances.}
{\small
\begin{minipage}{2\columnwidth}
\centering
\renewcommand{\footnoterule}{}
\label{table:results}
\begin{tabular}{c c c c c c c c c}
\hline\hline 
Source & $N(o-{\rm H_2D^+})$ & $x(o-{\rm H_2D^+})$ & $N({\rm C^{17}O})$ & $x({\rm C^{17}O})$ & $N({\rm N_2H^+})$ & $x({\rm N_2H^+})$ & $N({\rm N_2D^+})$ & $x({\rm N_2D^+})$ \\
       & [$10^{12}$ cm$^{-2}$] & [$10^{-11}$] & [$10^{14}$ cm$^{-2}$] & [$10^{-8}$] & [$10^{13}$ cm$^{-2}$] & [$10^{-10}$] & [$10^{12}$ cm$^{-2}$] & [$10^{-10}$]\\
\hline
IRAS 05399-0121 & $<1.4$ & $<2.3$ & $10.8\pm1.1$ & $2.7\pm0.4$ & $5.5\pm0.8$ & $10.7\pm2.0$ & $4.7\pm0.3$ & $1.1\pm0.1$ \\ [1ex] 
SMM 1 & $4.8\pm0.8$ & $11.5\pm2.3$ & $17.8\pm1.0$ & $7.4\pm0.9$ & $0.8\pm0.1$ & $3.1\pm0.6$ & $3.8\pm0.4$ & $1.6\pm0.2$ \\ [1ex] 
SMM 3 & $6.4\pm1.0$ & $5.9\pm1.1$ & $4.0\pm0.6$\tablefootmark{a} & $0.5\pm0.1$\tablefootmark{a} & $1.9\pm0.2$ & $1.6\pm0.3$ & $2.3\pm0.5$ & $0.3\pm0.1$ \\ [1ex] 
IRAS 05405-0117 & $<1.5$ & $<9.7$ & $3.2\pm0.3$ & $1.0\pm0.2$ & $0.5\pm0.1$ & $3.6\pm0.6$ & $0.3\pm0.2$ & $0.2\pm0.1$ \\ [1ex] 
SMM 6\tablefootmark{b} & $8.4\pm0.7$ & $30.2\pm5.6$ & $2.0\pm0.3$ & $0.9\pm0.2$ & $0.6\pm0.1$ & $2.4\pm0.4$ & $3.1\pm0.1$ & $1.3\pm0.2$ \\[1ex]       
            &   \ldots          &   \ldots     & $3.0\pm0.2$ & $1.3\pm0.2$ & $1.0\pm0.1$ & $3.8\pm0.7$ & $3.5\pm0.6$ & $1.5\pm0.3$ \\ [1ex] 

SMM 7 & $4.8\pm1.4$ & $11.5\pm3.7$ & $16.0\pm1.4$ & $5.9\pm0.9$ & $2.2\pm0.9$ & $6.8\pm2.9$ & $<3.9$\tablefootmark{c} & $<1.5$ \\ [1ex] 
\hline 
\end{tabular} 
\tablefoot{\tablefoottext{a}{The column density and fractional abundance of the C$^{18}$O isotopologue derived from observations towards the \textit{Spitzer} 24~$\mu$m peak position ($2\farcs2$ offset from the SABOCA peak) are $(7.8\pm1.0)\times10^{14}$ cm$^{-2}$ and $(9.6\pm1.6)\times10^{-9}$, respectively (\cite{miettinen2016}).}\tablefoottext{b}{The values on the first row for SMM~6 refer to the SABOCA 350~$\mu$m peak position (\cite{miettinen2013b}; their source SMM~6a), while those on the second row were derived towards our earlier target position (\cite{miettinenetal2012}).}\tablefoottext{c}{The N$_2$D$^+(3-2)$ line was not detected towards SMM~7 (\cite{miettinenetal2012}), and hence we estimated the column density upper limit from the $3\sigma$ line intensity upper limit and the N$_2$H$^+(3-2)$ linewidth divided by the average $\Delta v({\rm N_2H^+(3-2)})/\Delta v({\rm N_2D^+(3-2)})$ ratio ($=1.26$) in our sample. All the quoted upper limits take the uncertainties into account.}}
\end{minipage}}
\end{table*}

\subsection{CO depletion factor and the degree of deuterium fractionation}

To estimate the factors by which the CO molecules are
depleted in the target cores, we followed the analysis presented in Miettinen et al. (2012) with the modifications 
employed by Miettinen (2016). In short, the canonical or undepleted CO abundance was assumed to be $x({\rm CO})_{\rm can}=2.3\times10^{-4}$, 
and the $[^{16}{\rm O}]/[^{18}{\rm O}]$ and $[^{18}{\rm O}]/[^{17}{\rm O}]$ oxygen isotope ratios needed in the analysis were assumed to be 557 and 4.16, respectively (see \cite{miettinen2016} and references therein). The CO depletion factor was then calculated as $f_{\rm D}({\rm CO})=x({\rm CO})_{\rm can}/x({\rm CO})_{\rm obs}$, where the denominator is the observed CO abundance.

The degree of deuterium fractionation in N$_2$H$^+$ was calculated by dividing the column density of N$_2$D$^+$ by that of
N$_2$H$^+$. The values of $f_{\rm D}({\rm CO})$ and $[{\rm N_2D^+}]/[{\rm N_2H^+}]$ are tabulated in Table~\ref{table:properties}.

\subsection{CO depletion timescale}

The rate at which CO molecules freeze out onto dust grain surfaces, $k_{\rm fo}$, can be used to calculate the CO depletion timescale as (e.g. \cite{rawlings1992}; \cite{maret2013})

\begin{equation}
\tau_{\rm dep}=k_{\rm fo}^{-1}=\left(n_{\rm g}\sigma_{\rm g}v_{\rm CO}S_{\rm stick} \right)^{-1}\,,
\end{equation}
where $n_{\rm g}$ is the grain number density, $\sigma_{\rm g}=\pi a_{\rm g}^2$ is the 
mean geometric grain cross section with $a_{\rm g}$ being the average grain radius (assumed to be 0.1~$\mu$m), 
$v_{\rm CO}$ is the mean thermal speed of the
CO molecules, and $S_{\rm stick}$ is their sticking coefficient. The grain number density can be expressed as 
$n_{\rm g}=x_{\rm g}\times n({\rm H_2})$, where $x_{\rm g}$ is the fractional abundance of dust grains and $n({\rm H_2})$ is 
the H$_2$ number density. The value of $x_{\rm g}$ can be solved from the equality $x_{\rm g}\times m_{\rm g}=m_{\rm H_2}\times R_{\rm dg}$, where $m_{\rm g}$ is the mass of a dust grain and $m_{\rm H_2}$ is that of the H$_2$ molecule. Under the assumption of spherical dust grains, the mass of a grain is given by $m_{\rm g}=4/3\times\pi a_{\rm g}^3\rho_{\rm g}$, where $\rho_{\rm g}$ is the mass density 
of a grain (assumed to be 3~g~cm$^{-3}$). For a Maxwellian distribution of gas-phase CO molecules, the mean thermal speed is given by 
$v_{\rm CO}=\left(8/\pi \times k_{\rm B}T_{\rm kin}/m_{\rm CO} \right)^{1/2}$, where $m_{\rm CO}$ is the mass of the CO molecule. Finally, we assumed that $S_{\rm stick}=1$, which signifies that the CO molecules stick to the dust grains in each collision, which is expected to be a reasonable assumption at the low temperatures of the present target sources (\cite{burke1983}; \cite{bisschop2006}).

The volume-averaged H$_2$ number densities and the corresponding CO depletion timescales of the target sources are listed in Table~\ref{table:properties}. The former values were adopted from our previous studies of the Orion~B9 cores (\cite{miettinen2010}, 2012; \cite{miettinen2013b}), but they were scaled upward owing to the present assumptions about the source distance and dust properties. 

\subsection{Correlation plots}

\begin{figure*}
\begin{center}
\includegraphics[scale=0.33]{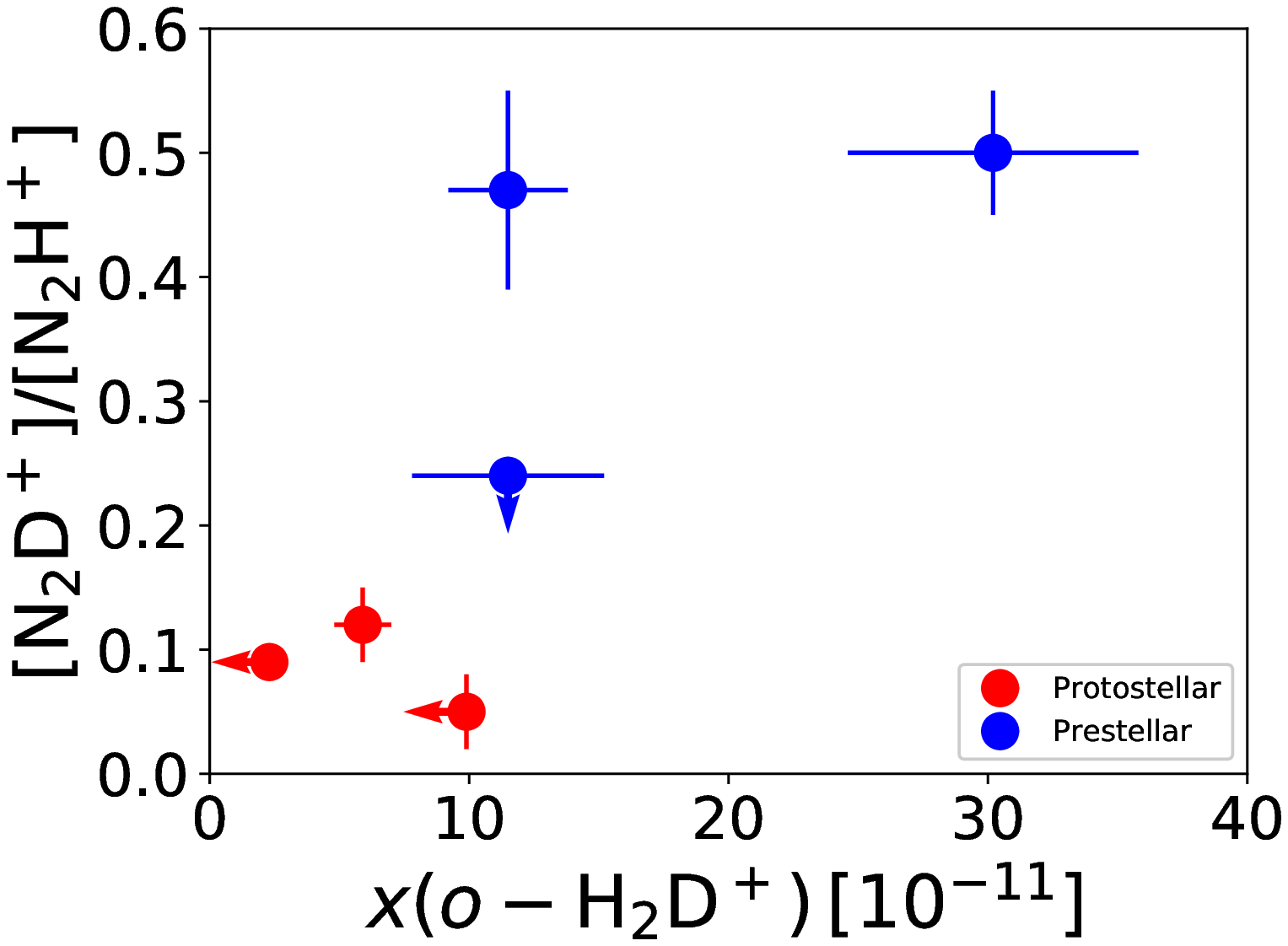}
\includegraphics[scale=0.33]{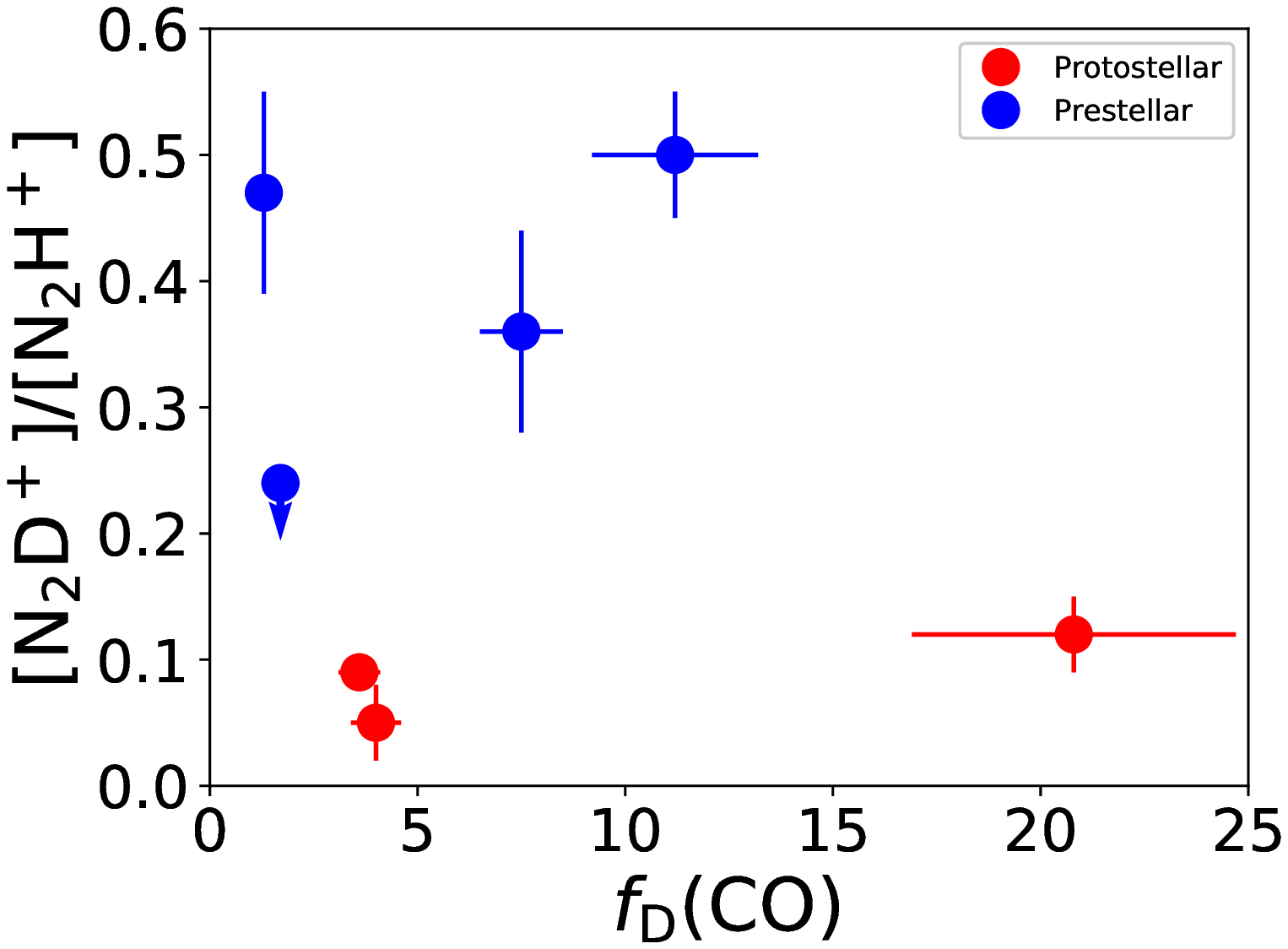}
\includegraphics[scale=0.33]{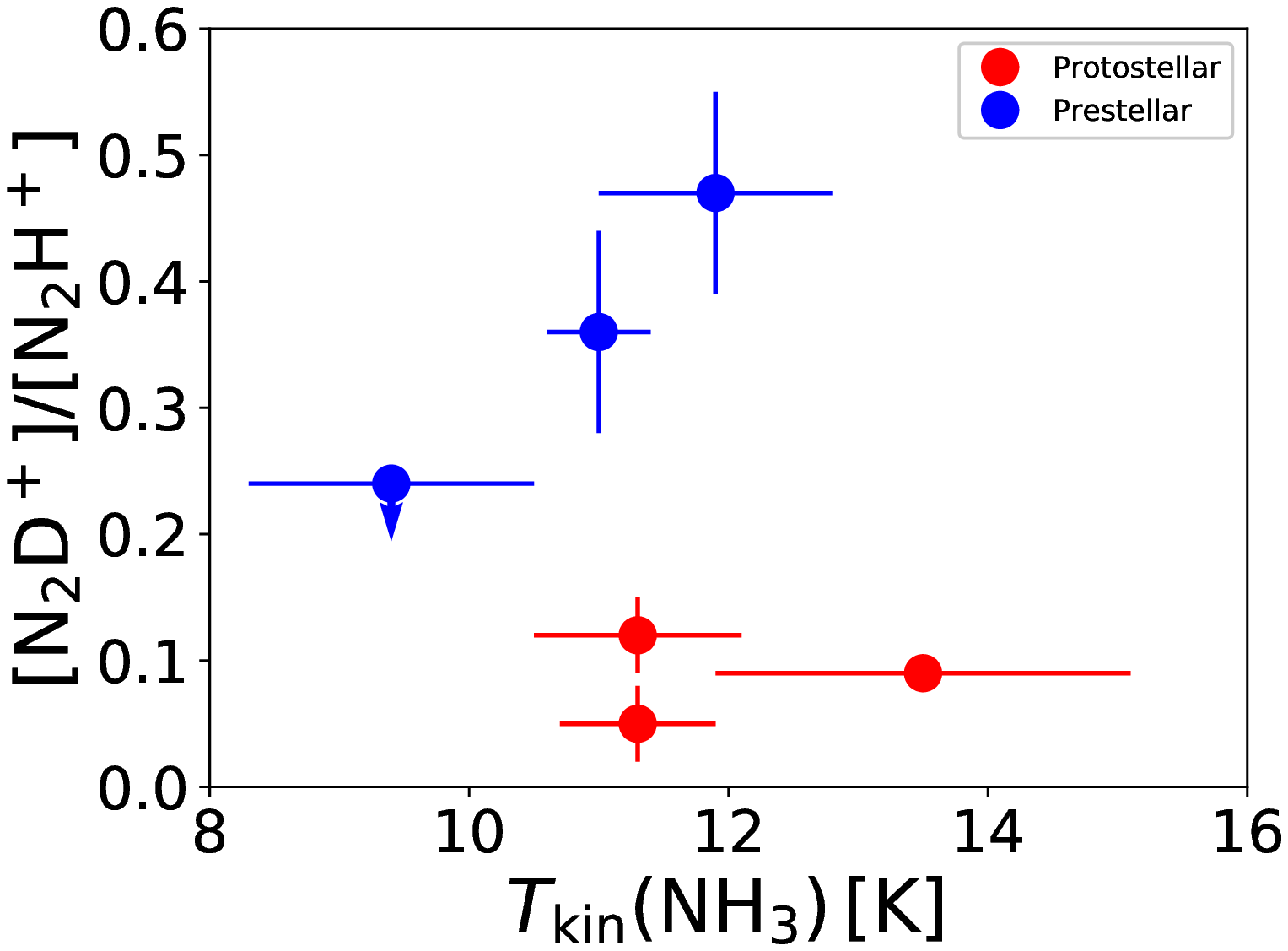}
\includegraphics[scale=0.33]{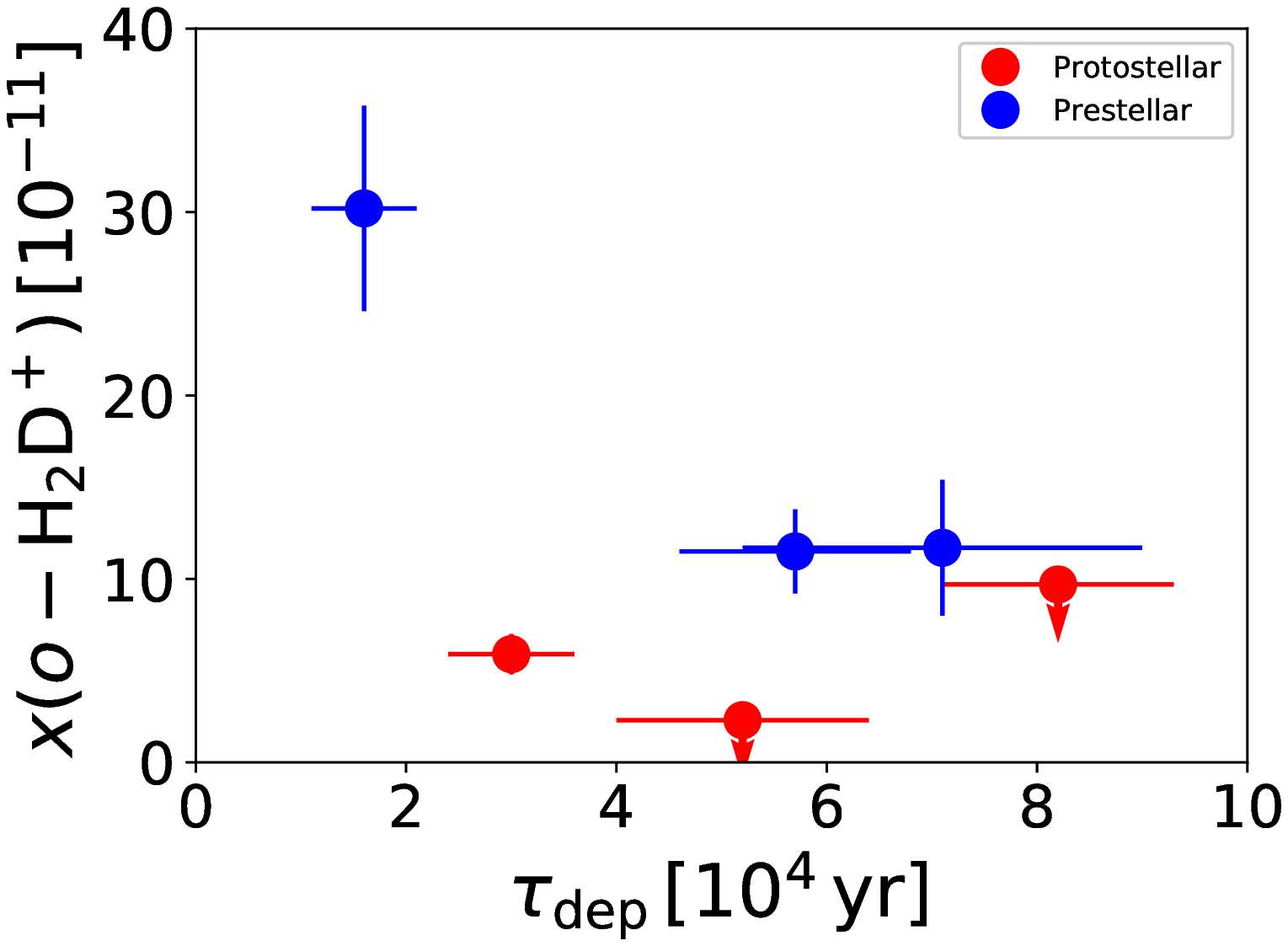}
\includegraphics[scale=0.33]{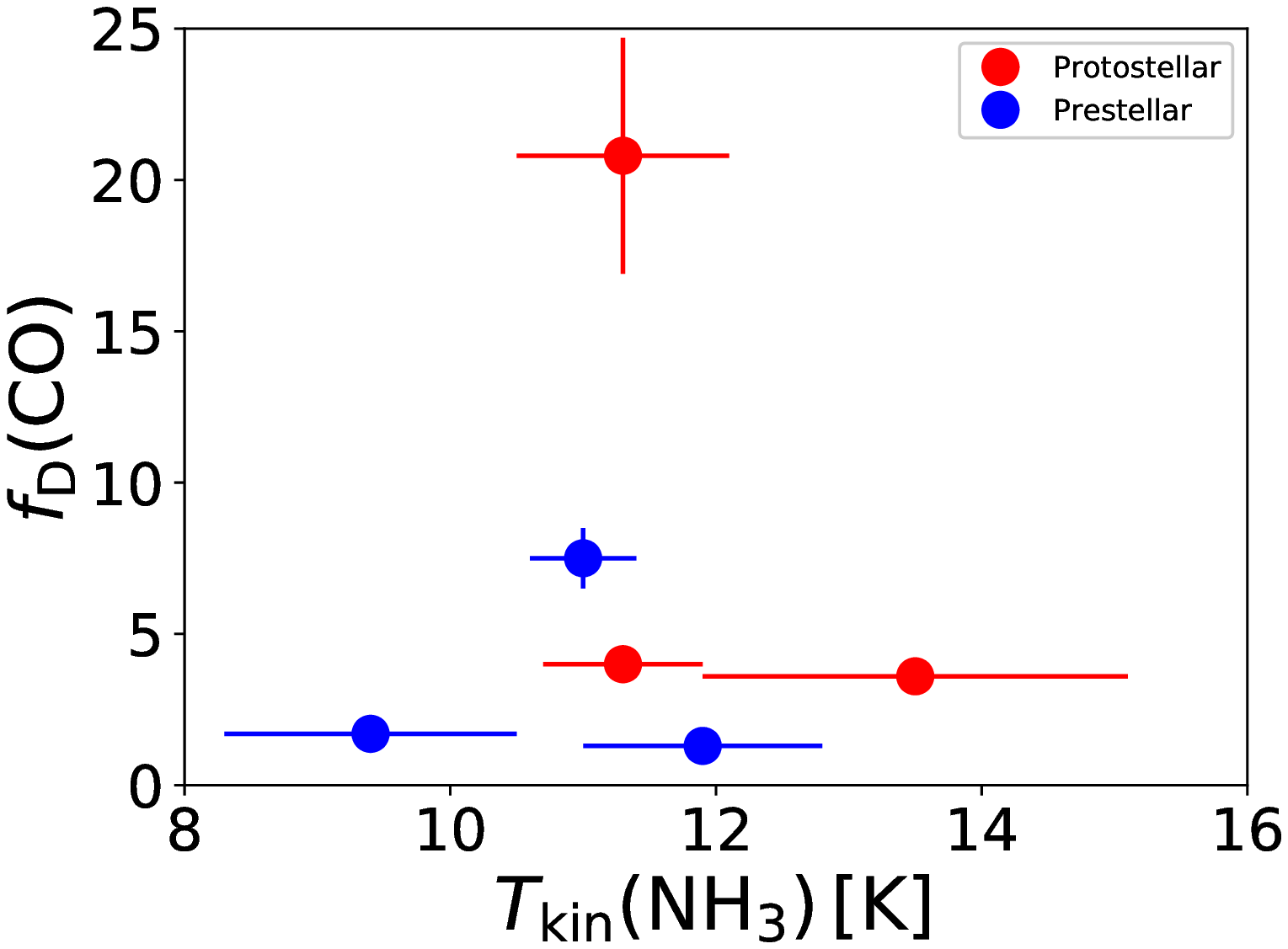}
\caption{(Top panels) Degree of deuterium fractionation ($[{\rm N_2D^+}]/[{\rm N_2H^+}]$) against \textit{o}-H$_2$D$^+$ abundance, CO depletion factor, and NH$_3$-based gas kinetic temperature. (Bottom panels) \textit{o}-H$_2$D$^+$ abundance against CO depletion timescale; CO depletion factor against gas temperature. The arrows pointing left and down indicate upper limits. In the top middle panel, the prestellar core SMM~6 has two data points that represent our two different target positions within the source (see Table~\ref{table:properties}).}
\label{figure:corr1}
\end{center}
\end{figure*}

To search for potential correlations between the different parameters derived in the present paper and our earlier studies, 
we made several scatter plots that are useful to visualise the relationship between two variables. Figure.~\ref{figure:corr1}
shows the degree of deuterium fractionation, or the $[{\rm N_2D^+}]/[{\rm N_2H^+}]$ ratio, as a function of the 
\textit{o}-H$_2$D$^+$ abundance, CO depletion factor, and the gas kinetic temperature derived from NH$_3$ by Miettinen et al. (2010).  The \textit{o}-H$_2$D$^+$ abundance is also shown as a function of the CO depletion timescale and the CO depletion factor as a function of the NH$_3$-based gas temperature. The prestellar and protostellar cores are shown in different colours (blue and red, respectively) to better illustrate how the corresponding data points populate the plotted parameter spaces.  

\begin{figure}[!htb]
\centering
\resizebox{0.98\hsize}{!}{\includegraphics{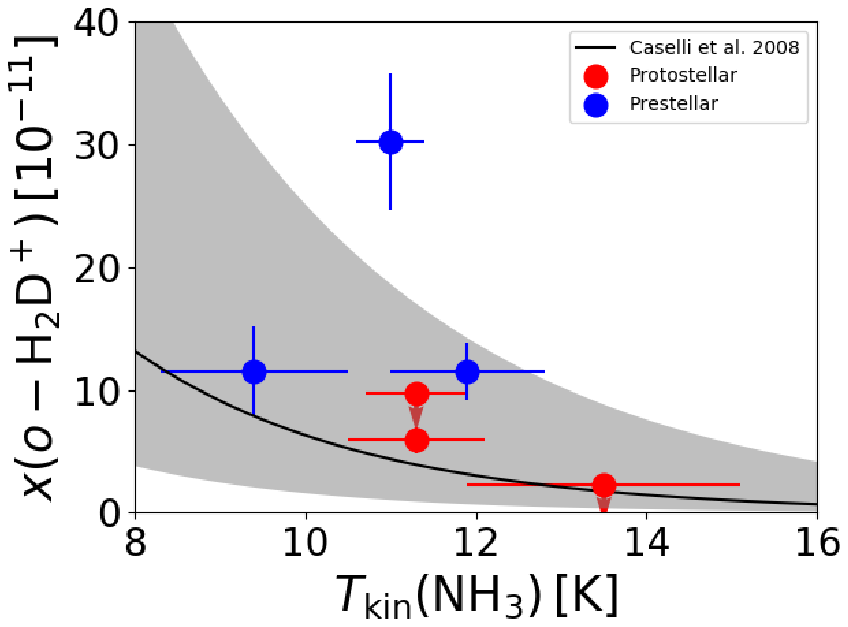}}
\caption{\textit{ortho}-H$_2$D$^+$ abundance as a function of the NH$_3$-based gas kinetic temperature. The down-pointing arrows indicate upper limits. The black-solid curve represents the $x(o-{\rm H_2D^+})-T_{\rm kin}$ relationship derived by Caselli et al. (2008), and the grey-shaded region shows its uncertainty.}
\label{figure:corr2}
\end{figure}

Figure~\ref{figure:corr2} shows the \textit{o}-H$_2$D$^+$ abundance as a function of the gas kinetic temperature. For comparison, 
the relationship between $x(o-{\rm H_2D^+})$ and $T_{\rm kin}$ derived by Caselli et al. (2008; their Eq.~(7)) is indicated 
in Fig.~\ref{figure:corr2}.

We remind the reader that the C$^{17}$O, NH$_3$, N$_2$H$^+$, and N$_2$D$^+$ observations used to construct the plots in Figs.~\ref{figure:corr1} and \ref{figure:corr2} were targeting the same positions. However, the new \textit{o}-H$_2$D$^+$ observations were made towards positions offset from our previous molecular line observations (the mean offset is $11\farcs9$ or 0.7 times the beam of the \textit{o}-H$_2$D$^+$ observations; Table~\ref{table:sources}), except for SMM~6 for which we also have C$^{17}$O, N$_2$H$^+$, and N$_2$D$^+$ data for the present target position (only $2\farcs2$ offset; \cite{miettinen2013b}).

\section{Discussion}

\subsection{Interrelations between \textit{o}-H$_2$D$^+$, CO depletion, and deuterium fractionation} 

Owing to the low deuteron affinity of the H$_2$D$^+$ molecule, it reacts eagerly with other species by donating a deuteron (e.g. \cite{roberts2002}). Therefore, deuterium enrichements are passed forward to heavier species. For example, one-third of the reactions between H$_2$D$^+$ and molecular nitrogen yield N$_2$D$^+$ (the remaining two-thirds lead to N$_2$H$^+$). The deuterium fractionation in a given species, such as N$_2$H$^+$ for example, is therefore expected to become higher for a higher abundance of H$_2$D$^+$. Despite the angular offset between the N$_2$H$^+$, N$_2$D$^+$, and \textit{o}-H$_2$D$^+$ observations (Sect.~3.5), this is qualitatively consistent with what we see in the top left panel in Fig.~\ref{figure:corr1}, that is the highest $[{\rm N_2D^+}]/[{\rm N_2H^+}]$ ratios are seen at the highest \textit{o}-H$_2$D$^+$ abundances (and vice versa). 

The H$_2$D$^+$ molecules are mainly destroyed in reactions with CO, which yield either HCO$^+$ or DCO$^+$. Therefore, the less gas-phase CO there is, the lower the destruction rate of H$_2$D$^+$ (e.g. \cite{roberts2000}). In other words, the higher the CO depletion factor, the higher the expected deuteration degree (e.g. \cite{vastel2006}). Although the data points plotted in the top middle panel in Fig.~\ref{figure:corr1} show a large scatter, the prestellar and protostellar core populations are each qualitatively consistent with the aforementioned description, that is the highest $[{\rm N_2D^+}]/[{\rm N_2H^+}]$ ratios are associated with the highest levels of CO depletion (see Sect.~4.2). 

Because the level of CO depletion via freeze-out becomes higher at higher gas densities (e.g. \cite{bacmann2002}), one would also expect higher $f_{\rm D}({\rm CO})$ values, and hence higher deuteration levels at lower gas temperatures (e.g. \cite{emprechtinger2009}). These trends are not visible in the top right and bottom right panels of Fig.~\ref{figure:corr1}. If anything, there appears to be a positive correlation between $[{\rm N_2D^+}]/[{\rm N_2H^+}]$ and $T_{\rm kin}$ among the prestellar cores. However, the gas temperature range explored here is fairly narrow, from $9.4\pm1.1$~K to $13.5\pm1.6$~K, which might render the trend imperceptible. Another potential issue is the fact that the gas temperatures plotted in Fig.~\ref{figure:corr1} were derived from the $(1,\,1)$ and $(2,\,2)$ inversion lines of \textit{para}-NH$_3$, which have a much lower critical density ($\sim10^3$~cm$^{-3}$) than the $J=3-2$ transitions of N$_2$H$^+$ and N$_2$D$^+$ ($\sim10^6$~cm$^{-3}$) from which the deuteration level was derived (\cite{miettinen2010}, 2012; \cite{miettinen2016}). Therefore, the derived $[{\rm N_2D^+}]/[{\rm N_2H^+}]$ ratios and $T_{\rm kin}({\rm NH_3})$ values might refer to different gas components, which would explain the lack of the expected negative correlation in the top right panel of Fig.~\ref{figure:corr1}. 

\begin{table*}
\renewcommand{\footnoterule}{}
\caption{Volume-averaged H$_2$ number density, CO depletion factor and timescale, and degree of deuteration of the target sources.}
{\normalsize
\begin{minipage}{2\columnwidth}
\centering
\label{table:properties}
\begin{tabular}{c c c c c}
\hline\hline 
Source & $\langle n({\rm H}_2) \rangle$ & $f_{\rm D}({\rm CO})$ & $\tau_{\rm dep}$ & $[{\rm N_2D^+}]/[{\rm N_2H^+}]$ \\ 
       & [$10^4$~cm$^{-3}$]  &   & [$10^4$~yr]  & \\ 
\hline 
IRAS 05399-0121 & $10.2\pm2.4$ & $3.6\pm0.5$ & $5.2\pm1.2$ & $0.09\pm0.01$ \\
SMM 1 & $9.9\pm1.7$ & $1.3\pm0.2$ & $5.7\pm1.0$ & $0.47\pm0.08$ \\
SMM 3\tablefootmark{a} & \ldots & $42.7\pm7.0$ & \ldots & \ldots \\
      & $19.5\pm3.9$ & $20.8\pm3.9$ & $3.0\pm0.6$ & $0.12\pm0.03$ \\
IRAS 05405-0117 & $7.1\pm0.9$ & $4.0\pm0.6$ & $8.2\pm1.1$ & $0.05\pm0.03$ \\
SMM 6\tablefootmark{b} & $36.4\pm11.8$ & $11.2\pm2.0$ & $1.6\pm0.5$ & $0.50\pm0.05$ \\
                       & $4.1\pm0.6$ & $7.5\pm1.0$ & $14.3\pm2.0$ & $0.36\pm0.08$ \\
SMM 7 & $8.9\pm2.4$ & $1.7\pm0.2$ & $7.1\pm1.9$ & $<0.24$ \\
\hline
\end{tabular} 
\tablefoot{\tablefoottext{a}{The CO depletion factor quoted on the first row was derived from the C$^{18}$O observations towards the \textit{Spitzer} 24~$\mu$m peak position of the source, which lies $2\farcs2$ to the north-west of the SABOCA peak (\cite{miettinen2016}), while the values on the second row were derived towards our earlier target position (\cite{miettinenetal2012}).}\tablefoottext{b}{The values on the first row for SMM~6 refer to the SABOCA 350~$\mu$m peak position (\cite{miettinen2013b}; their source SMM~6a), while those on the second row were derived towards our earlier target position (\cite{miettinenetal2012}).}}
\end{minipage} }
\end{table*}

\subsection{Comparison of the H$_2$D$^+$-based deuteration between the prestellar and protostellar cores in Orion~B9}

As illustrated in the top left panel of Fig.~\ref{figure:corr1}, the studied prestellar cores show higher \textit{o}-H$_2$D$^+$ abundances and $[{\rm N_2D^+}]/[{\rm N_2H^+}]$ deuteration ratios than the protostellar cores. This can be understood by considering the feedback effects from the protostar(s) deeply embedded in the protostellar cores. The central protostar increases its mass by accreting gas from the surrounding envelope via a circumstellar disk, and this process is associated with both heating of the surrounding medium and protostellar jets that drive molecular outflows. When the gas temperature exceeds $\sim25$~K, reaction~(\ref{eqn:deut}) starts to operate from right to left, which leads to the decrease of deuteration (e.g. \cite{vastel2006}). At the same time, thanks to both central heating and outflow shocks, CO molecules can be released from the icy grain mantles back into the gas phase, which leads to the destruction of H$_2$D$^+$.

That the prestellar cores populate a higher $[{\rm N_2D^+}]/[{\rm N_2H^+}]$ regime than the more evolved protostellar cores is also 
visible in the top middle and right panels in Fig.~\ref{figure:corr1}. Interestingly, the lowest CO depletion factor in our sample was derived for the prestellar core SMM~1, namely $f_{\rm D}({\rm CO})=1.3\pm0.2$, which is consistent with no depletion. On the other hand, the protostellar core SMM~3 exhibits the highest level of CO depletion, $f_{\rm D}({\rm CO})=42.7\pm7.0$, towards the core centre and $f_{\rm D}({\rm CO})=20.8\pm3.9$ towards the envelope ($16\farcs7$ or 0.034~pc projected separation). The Class~0 object SMM~3 is the strongest source of 350~$\mu$m and 870~$\mu$m emission in Orion~B9 (\cite{miettinen2009}, 2012), and is therefore likely to be in a very early stage of protostellar evolution, where the dusty envelope is still resembling the properties that prevailed at its prestellar phase. The prestellar core SMM~1 on the other hand might be affected by the northwest-southeast oriented outflow driven by the nearby Class~0/I protostar IRAS~05399 (see \cite{miettinen2013a} and references therein). This could increase the CO abundance in the gas phase in SMM~1 via outflow shock desorption, or simply by pushing extra CO gas towards SMM~1 that is being captured in our observations. We note that the outflow motions from IRAS~05399 are also indicated by the red asymmetric profile of the DCO$^+(5-4)$ line detected in the present work (see Fig.~\ref{figure:otherspectra2}). Nevertheless, the low CO depletion factor in SMM~1 is still puzzling in terms of the high deuteration observed in the source ($[{\rm N_2D^+}]/[{\rm N_2H^+}]\sim0.5$). Perhaps SMM~1 is seen in a specific stage, where the abundant CO molecules have not yet had time to bring down the deuteration in N$_2$H$^+$. 

\subsection{Orion~B9 in a wider context of H$_2$D$^+$ studies of star-forming regions}

As the main comparison sample of \textit{o}-H$_2$D$^+$ observations towards dense cores, we used the 
Caltech Submillimeter Observatory (CSO) survey by Caselli et al. (2008). To our knowledge, the Caselli et al. (2008) study is the largest H$_2$D$^+$ survey published so far, and therefore provides a useful comparison sample with the present results. These authors observed the same \textit{o}-H$_2$D$^+(1_{1,\,0}-1_{1,\,1})$ transition as we did towards ten starless cores and six 
protostellar cores. The line was detected in seven ($70\%$) and four ($\sim67\%$) of these targets, respectively. 
We detected \textit{o}-H$_2$D$^+$ in all three of our prestellar cores, but in only one of the three protostellar cores ($33\%$ detection rate). We note that the Caselli et al. (2008) sample included a target position in Orion~B9, which was called Ori~B9, but that position is not coincident with any significant LABOCA dust emission (see Fig.~\ref{figure:map}), which probably explains their non-detection. As discussed in Sect.~1, the latter position is close to one of the positions observed in \textit{o}-H$_2$D$^+$ by Harju et al. (2006; $10\farcs4$ offset from their source Ori~B9~N). We also note that Ori~B9 is often considered to be a massive dense core in the literature (e.g. \cite{harju2006}; \cite{caselli2008}; \cite{pillai2012}), but the source is composed of a system of low-mass dense cores. 

Caselli et al. (2008) used a somewhat different method to calculate the \textit{o}-H$_2$D$^+$ column densities (e.g. their calculation made use of the photon escape probability), and therefore it is important to decipher whether or not the two methods lead to similar results. For this purpose, we took the line parameters for the two strongest \textit{o}-H$_2$D$^+$ detections by Caselli et al. (2008), namely those towards L~1544 and L~183 (their Tables~2 and 3; the values valid at a critical density of $10^5$~cm$^{-3}$), and derived $N(o-{\rm H_2D^+})$ values that agree within factors of $1.1\pm0.1$ (our nominal values being $10\%$ higher). We therefore conclude that the two methods yield results that are in good agreement with each other, and a direct comparison is reasonable. 

The \textit{o}-H$_2$D$^+$ line widths derived by Caselli et al. (2008) for their starless and protostellar cores are $0.33-0.73$~km~s$^{-1}$ and $0.51-1.49$~km~s$^{-1}$ (the authors adopted a line width of 0.42~km~s$^{-1}$ for Ori~B9 from Harju et al. (2006), but that value refers to the starless source Ori~B9~N, and is hence neglected from the latter range). The corresponding averages are 0.47~km~s$^{-1}$ and 0.93~km~s$^{-1}$. We found a factor 1.34 broader average \textit{o}-H$_2$D$^+$ line width (0.63~km~s$^{-1}$) for our prestellar cores, which could indicate that Orion~B9 is dynamic in nature (see Sect.~1). On the other hand, the line width for our only protostellar source detection (SMM~3; 0.74~km~s$^{-1}$) is comparable with those in the prestellar objects (only 1.17 times broader than the average line width for prestellar cores), which is at odds with the average trend in the Caselli et al. (2008) sample, but supports the aforementioned hypothesis that SMM~3 is in its very early Class~0 stage. 

The \textit{o}-H$_2$D$^+$ column densities derived for starless cores by Caselli et al. (2008) are $(0.2-4.1)\times10^{13}$~cm$^{-2}$, while those for protostellar cores are $(2-9)\times10^{12}$~cm$^{-2}$. The corresponding fractional abundances are $(2.4-34.1)\times10^{-11}$ and $(0.6-5.0)\times10^{-11}$. The $N(o-{\rm H_2D^+})$ values for the Orion~B9 prestellar cores are derived to be $\sim(5-8)\times10^{12}$~cm$^{-2}$, with the corresponding fractional abundances being $\sim(12-30)\times10^{-11}$. Although the former values resemble the lower end values found by Caselli et al. (2008), our $x(o-{\rm H_2D^+})$ values are closer to the highest values in the CSO survey. Although the \textit{o}-H$_2$D$^+$ column density we derived for 
SMM~3 is bracketed by the range derived by Caselli et al. (2008) for their dense cores associated with ongoing star formation, the $x(o-{\rm H_2D^+})$ value of the source ($6\times10^{-11}$) appears to be higher. However, derivation of the fractional abundances is susceptible to the assumptions used to calculate the H$_2$ column density, and therefore it might be more reliable to compare the $N(o-{\rm H_2D^+})$ values.

The only significant correlation between core properties found by Caselli et al. (2008) was that between $x(o-{\rm H_2D^+})$ and $T_{\rm kin}$ (see our Fig.~\ref{figure:corr2}). Within the uncertainties, our results appear to be consistent with this relationship with a notable exception being the prestellar core SMM~6. However, a notable caveat in our analysis is that our $T_{\rm kin}$ values were derived from ammonia observations towards positions that are different from the present target positions (i.e. $6\arcsec-18\farcs7$
offsets from the SABOCA 350~$\mu$m peak positions; Table~\ref{table:sources}), although the large beam size of the former data ($40\arcsec$ HPBW) encompasses the \textit{o}-H$_2$D$^+$ targets. Nevertheless, despite this weakness in our methodology, at least half of our sources are consistent even with the nominal $x(o-{\rm H_2D^+})-T_{\rm kin}$ correlation derived by Caselli et al. (2008). As discussed above, the values of $x(o-{\rm H_2D^+})$ are sensitive to the way the corresponding $N({\rm H_2})$ are calculated, and this can also affect the comparison made here. 

Regarding the validity of using the NH$_3$-based gas temperature to compare with the \textit{o}-H$_2$D$^+$ data, we also estimated 
the value of $T_{\rm kin}$ for SMM~6 from the observed line widths of \textit{o}-H$_2$D$^+$ and N$_2$D$^+(3-2)$; see \cite{miettinen2013b}. If the observed transitions of these two species originate in a common gas component, they should be sensitive to the same gas motions and gas temperature, which in turn would imply equally large non-thermal velocity dispersions (see \cite{friesen2010}; Eq.~(7) therein). From the equality $\sigma_{\rm NT}(o-{\rm H_2D^+})=\sigma_{\rm NT}({\rm N_2D^+})$ we derived a value of $T_{\rm kin}=12.2\pm 1.2$~K for SMM~6, which agrees within the uncertainties with the value $T_{\rm kin}({\rm NH_3})=11.0\pm0.4$~K derived by Miettinen et al. (2010). For comparison, using the N$_2$H$^+(4-3)$ line width in the analysis (see Appendix~A), the temperature estimate becomes higher, namely $16.9\pm 2.7$~K. The former comparison is expected to be more accurate because N$_2$D$^+$ is formed from H$_2$D$^+$ (cf. \cite{friesen2010}), and supports the use of $T_{\rm kin}({\rm NH_3})$ values in our analysis.
 
Caselli et al. (2008) suggested the following reasons for the negative $x(o-{\rm H_2D^+})-T_{\rm kin}$ relationship 
visible in their sample. First, the warmer the source, the smaller the \textit{o}-H$_2$D$^+$-emitting region might be, which 
would lead to a beam dilution effect (the line intensity is diluted by the ratio of the solid angle subtended by the emission region to the beam solid angle). Secondly, the warmest sources in the Caselli et al. (2008) sample were generally identified as being
the most distant ones ($>300$~pc), which could also be an issue owing to beam dilution. Thirdly, the physical and chemical interpretation is that the warmer the gas, the lower the degree of CO depletion, and hence the higher the destruction rate of H$_2$D$^+$. The authors also pointed out that the heating by a central protostar might not significantly increase the gas-phase CO abundance (as supported by the bottom right panel in our Fig.~\ref{figure:corr1}), but that the process might instead affect the \textit{ortho}-\textit{para} ratio of H$_2$D$^{+}$ in such a way that the \textit{o}-H$_2$D$^+$ abundance decreases (see e.g. \cite{sipila2017} for the \textit{ortho}-\textit{para} conversion reactions of H$_2$D$^+$). 

Pillai et al. (2012) derived $N(o-{\rm H_2D^+})$ values of $(1.9-3.8)\times10^{12}$~cm$^{-2}$ and $x(o-{\rm H_2D^+})$ values 
of $(0.9-2.1)\times10^{-11}$ towards five sources in the DR~21 filament of Cygnus X, which is a high-mass star-forming region. 
The authors assumed a $T_{\rm ex}$ value of 10~K for the whole sample and different dust properties from those assumed here
to calculate the H$_2$ column density (and hence $x(o-{\rm H_2D^+})$). Taking into account the differences in the dust opacity ($\kappa_{\nu}\propto \nu^{-\beta}$, where the dust emissivity index is $\beta\simeq 1.8$ for our adopted dust model (see \cite{miettinen2013a} and references therein)) and gas-to-dust mass ratio (we used a factor of 1.41 higher value), the fractional abundances reported by Pillai et al. (2012) should be scaled down by a factor of 0.47 for a more meaningful comparison. Our \textit{o}-H$_2$D$^+$ column densities in the detected sources appear to be somewhat higher than in DR~21, but the upper limits for the protostellar cores IRAS~05399 and IRAS~05405 are less than the lowest value found by Pillai et al. (2012). Also, the \textit{o}-H$_2$D$^+$ abundances we derived appear to be higher than in DR~21 (e.g. a factor of 30 difference between the highest values). 

Giannetti et al. (2019) reported $N(o-{\rm H_2D^+})$ values of $(<2.6-33.3)\times10^{11}$~cm$^{-2}$ towards three clumps 
in the G351.77-0.51 filament that have the potential to form high-mass stars. These column densities also appear to be lower than
in our sample. Hence, the \textit{o}-H$_2$D$^+$ abundances appear to be higher in low-mass star-forming regions than in massive star-forming regions. The process of low-mass star formation is characterised by at least an order of magnitude longer (and colder) starless core phase compared to that of high-mass stars (e.g. \cite{brunken2014}; \cite{tige2017}), which supports the development of matured deuterium chemistry in low-mass dense cores. 

\subsection{Relative abundances of \textit{o}-H$_2$D$^+$, N$_2$D$^+$, and DCO$^+$ as evolutionary indicators}

Giannetti et al. (2019) found that as the clump evolves, the \textit{o}-H$_2$D$^+$ abundance drops while that of N$_2$D$^+$ behaves in the opposite way, and hence the [\textit{o}-${\rm H_2D^+}]/[{\rm N_2D^+}]$ ratio decreases as a function of evolution (see their Fig.~3). The authors suggested that this trend could be caused by the conversion of H$_2$D$^+$ to the doubly and triply deuterated species D$_2$H$^+$ and D$_3^+$ (${\rm H_2D^+}+{\rm HD}\rightarrow {\rm D_2H^+}$ and ${\rm D_2H^+}+{\rm HD}\rightarrow {\rm D_3^+}$), associated with the formation of N$_2$D$^+$ in the reactions ${\rm D_2H^+}+{\rm N_2}\rightarrow {\rm N_2D^+}$ and ${\rm D_3^+}+{\rm N_2}\rightarrow {\rm N_2D^+}$ (e.g. \cite{pagani2009a}, 2011). In the protostellar stage, where the gas-phase CO abundance is expected to be higher than in the starless phase (at least near the central protostar), the H$_2$D$^+$ and N$_2$D$^+$ molecules are being destroyed by CO. Also, the reverse reaction~(\ref{eqn:deut}) becomes relevant at higher temperatures, which leads to the drop of the [\textit{o}-${\rm H_2D^+}]/[{\rm N_2D^+}]$ ratio. To examine if the sources in our sample exhibit such a trend, we first estimated the relative evolutionary stages of the target prestellar cores using the CO depletion timescales derived in Sect.~3.4. The chemical CO depletion timescale can be interpreted as a lower limit to the age of the core (\cite{maret2013}), and the values given in Table~\ref{table:properties} suggest that SMM~6 could be the youngest prestellar core in our sample (age $>10^4$~yr), followed by SMM~1 ($>4.7 \times 10^4$~yr) and SMM~7 ($>5.2 \times 10^4$~yr). We note that in the present study we observed the densest condensation in SMM~6 (\cite{miettinen2013b}; their source SMM~6a), while the CO depletion timescale derived 
for the lower density elongated parent core suggests an age of $>1.2 \times 10^5$~yr. 

Miettinen et al. (2009, 2012) studied the spectral energy distributions (SEDs) of the Orion~B9 protostellar cores (see also \cite{miettinen2013a}; \cite{miettinen2016}). As part of the SED analysis, the authors derived the $M_{\rm tot}/L_{\rm bol}^{0.6}$ ratios of the sources, where $M_{\rm tot}$ is the total (gas plus dust) mass of the core (essentially the envelope mass) and $L_{\rm bol}$ is the bolometric luminosity. This ratio is found to decrease with time (it is related to the weakening of the protostellar outflow strength), and can therefore be used as an evolutionary indicator (\cite{bontemps1996}). On the basis of this analysis, SMM~3 is the youngest protostellar core in our sample, followed by IRAS~05405 and then IRAS~05399. Indeed, in our earlier studies we found that SMM~3 and IRAS~05405 are Class~0 objects, while IRAS~05399 appears to be in the transition phase from Class~0 to I. 

\begin{figure*}
\begin{center}
\includegraphics[scale=0.5]{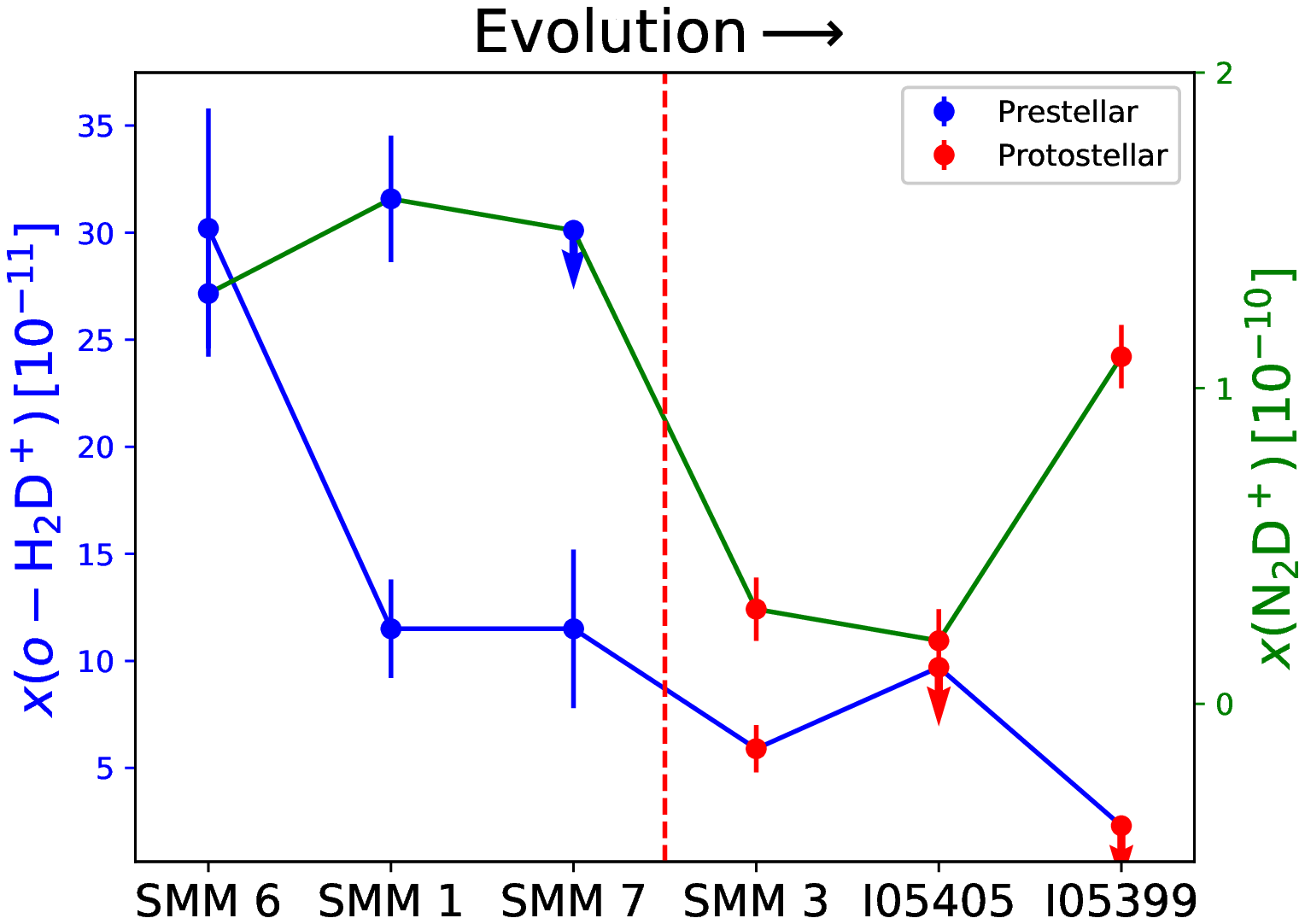}
\includegraphics[scale=0.5]{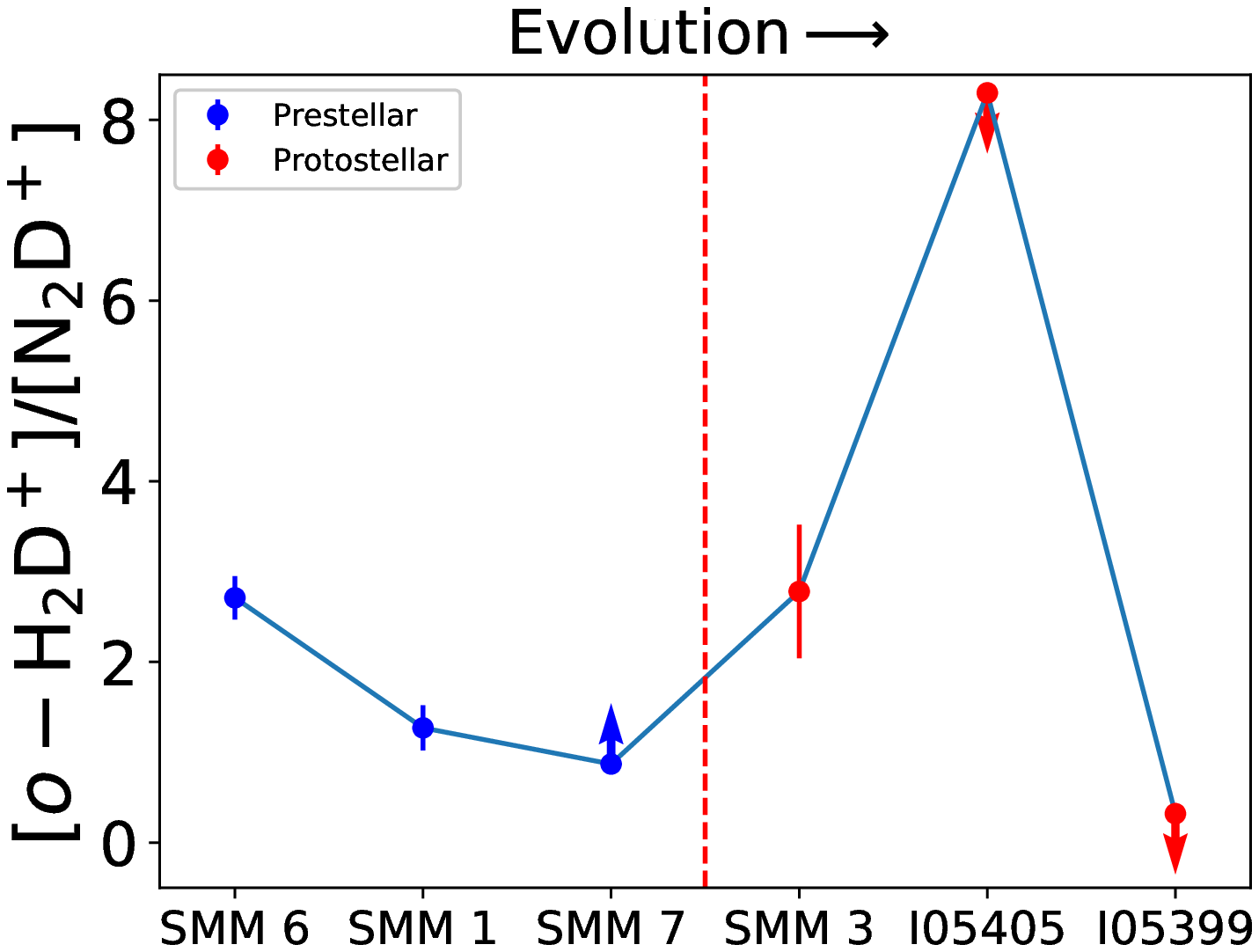}
\caption{Fractional abundances of \textit{o}-H$_2$D$^+$ and N$_2$D$^+$ (\textit{left panel}) and their abundance ratio (\textit{right panel}) plotted as a function of the potential source evolutionary sequence (see text for details). The arrows pointing up and down indicate the lower and upper limits, respectively. The red, vertical dashed line separates the prestellar phase of evolution on the left from the protostellar phase on the right.}
\label{figure:evolution1}
\end{center}
\end{figure*}

Under the assumption of the aforementioned evolutionary sequence, in Fig.~\ref{figure:evolution1} we plot the fractional abundances of \textit{o}-H$_2$D$^+$ and N$_2$D$^+$ and the [\textit{o}-${\rm H_2D^+}]/[{\rm N_2D^+}]$ ratio (calculated from the column densities) as a function of source evolution. Indeed, the \textit{o}-H$_2$D$^+$ abundance appears to drop as the core evolves in agreement with the finding by Giannetti et al. (2019). We note that the \textit{o}-H$_2$D$^+$ abundances in SMM~1 and SMM~7 are very similar to each other (their ratio is $1.0\pm0.4$), but so are their CO depletion times (with a ratio of $0.8\pm0.3$), which suggests that the cores are in comparable stages of evolution (cf.~the bottom left panel in Fig.~\ref{figure:corr1}). However, the behaviour of the N$_2$D$^+$ abundance is more fluctuating, which might be a manifestation of the fact that our N$_2$D$^+$ observations were offset from the core centres (except for SMM~6). 

As a result of the aforementioned fluctuation in the observed N$_2$D$^+$ abundance, the relative abundance of \textit{o}-H$_2$D$^+$ and N$_2$D$^+$ does not exhibit a clear decreasing trend with core evolution as found by Giannetti et al. (2019). However, owing to the censored values (especially the upper [\textit{o}-${\rm H_2D^+}$]/$[{\rm N_2D^+}]$ limits for IRAS~05405 and IRAS~05399), the existence of a decreasing trend is still possible, although SMM~3 would still be an outlier from such trend. 

\begin{figure*}
\begin{center}
\includegraphics[scale=0.5]{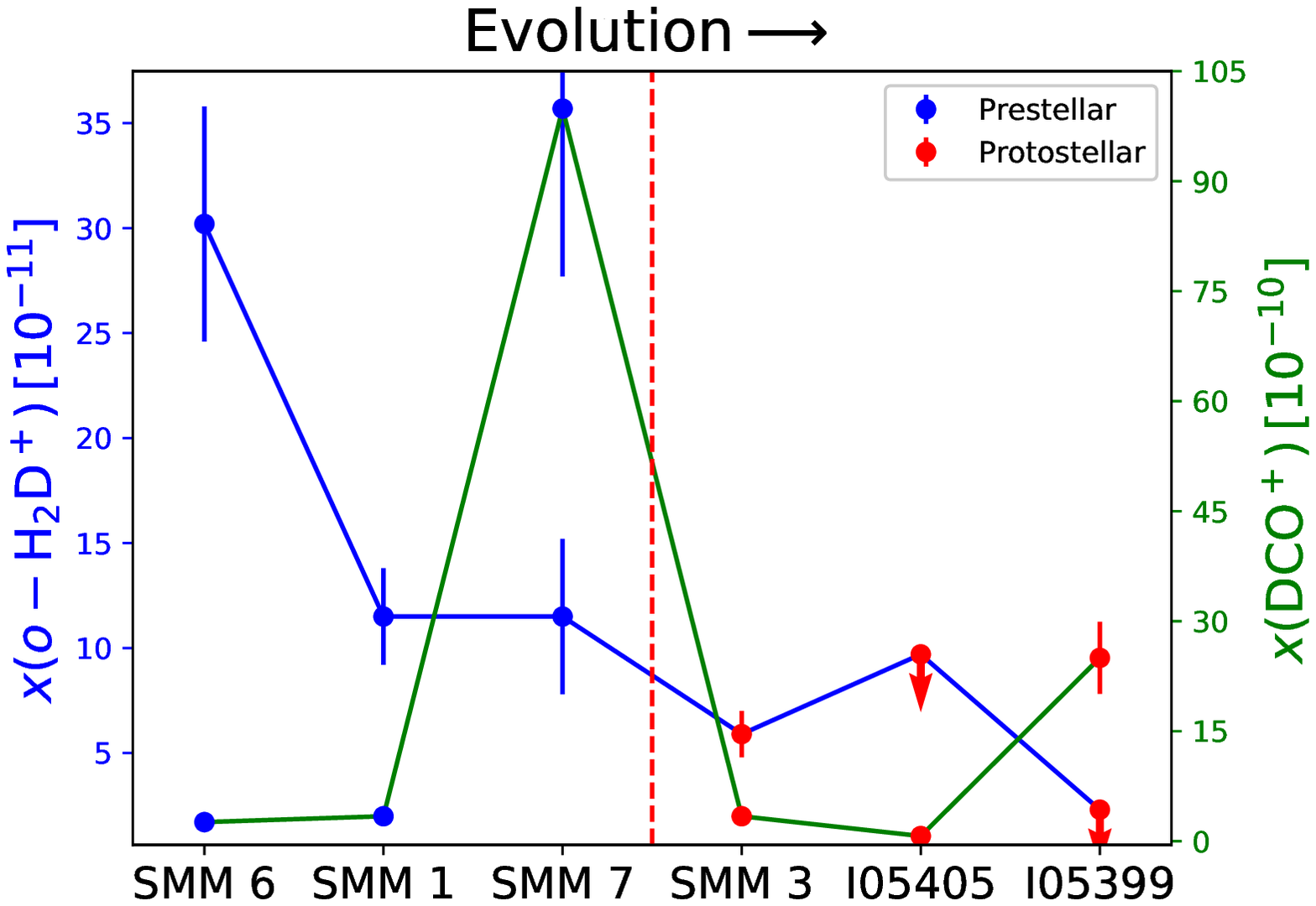}
\includegraphics[scale=0.5]{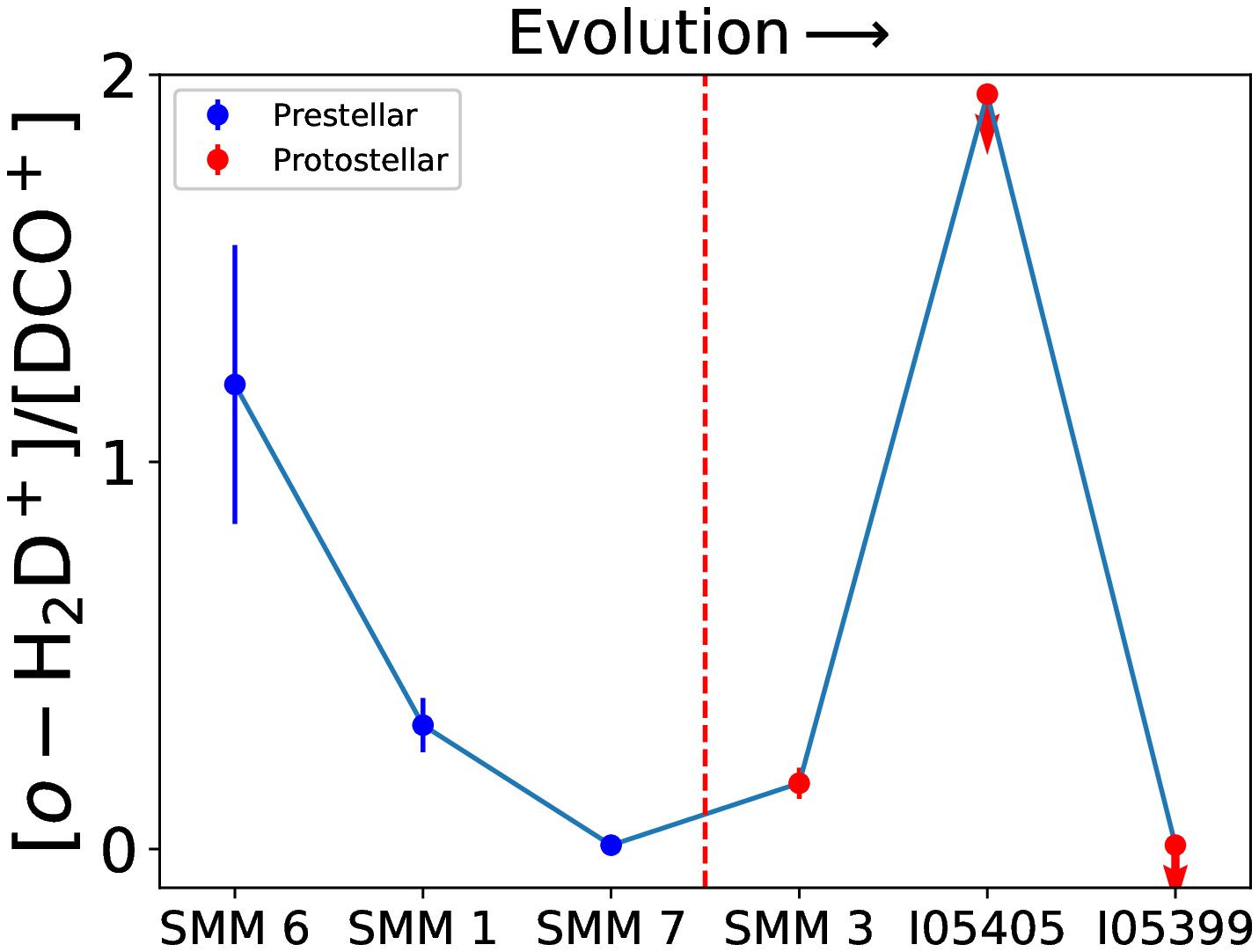}
\caption{Same as Fig.~\ref{figure:evolution1} but using DCO$^+$ in the analysis.}
\label{figure:evolution2}
\end{center}
\end{figure*}

Figure~\ref{figure:evolution2} is similar to Fig.~\ref{figure:evolution1} but DCO$^+$ was used in the analysis instead of N$_2$D$^+$ (see Appendix~A). The DCO$^+(5-4)$ data were observed from the exact same positions as the \textit{o}-H$_2$D$^+$ data (and simultaneosly in the observed frequency bands). Another benefit of using DCO$^+$ in the analysis is that the species was detected towards all six target sources. The DCO$^+$ abundance does not appear to exhibit any evolutionary trend, and the [\textit{o}-${\rm H_2D^+}]/[{\rm DCO^+}]$ ratio behaves in a similar way as the [\textit{o}-${\rm H_2D^+}]/[{\rm N_2D^+}]$ ratio. However, the prestellar regime shows a somewhat stronger hint of a decreasing trend than in the case of [\textit{o}-${\rm H_2D^+}]/[{\rm N_2D^+}]$. If the decrease of the \textit{o}-H$_2$D$^+$ abundance is indeed driven by its conversion to D$_2$H$^+$ and D$_3^+$, then DCO$^+$ can form in the reactions ${\rm D_2H^+}+{\rm CO}\rightarrow {\rm DCO^+}$ and ${\rm D_3^+}+{\rm CO}\rightarrow {\rm DCO^+}$ (e.g. \cite{pagani2009a}, 2011). On the other hand, the DCO$^+$ formation via these reactions is confronted by the abundance of gas-phase CO, which is first expected to decrease as the core evolves in its prestellar stage, and then to increase in the protostellar stage owing to desorption of icy grain mantles. The difference compared to the case of N$_2$D$^+$ above is that even if the gas-phase CO abundance is very low owing to its depletion, the gas can still be rich in N$_2$. For example, N$_2$ has about 18\% lower binding energy than CO, and therefore N$_2$ can evaporate from dust grain mantles more quickly (see \cite{giannetti2019} and references therein). Another chemical pathway that can contribute to the formation of DCO$^+$ is the reaction 
${\rm CH_3^+}+{\rm HD}\rightarrow {\rm CH_2D^+}$, followed by the reaction ${\rm CH_2D^+}+{\rm O}\rightarrow {\rm DCO^+}$ (e.g. \cite{favre2015}; \cite{salinas2017}). However, because the latter reactions are relevant at warm temperatures of $T>50$~K, they are expected 
to play a role in the formation of DCO$^+$ around the hot corinos of low-mass protostellar cores, where the temperature is $\sim100$~K (see \cite{ceccarelli2007} for a review).

To summarise, although we found that the \textit{o}-H$_2$D$^+$ abundance drops as the core evolves, the reliability of using of the [\textit{o}-${\rm H_2D^+}]/[{\rm N_2D^+}]$ and [\textit{o}-${\rm H_2D^+}]/[{\rm DCO^+}]$ abundance ratios as evolutionary indicators remains inconclusive. Larger source samples and observations of additional deuterated species like D$_2$H$^+$ would be particularly useful to better understand the temporal behaviour of the abundances of H$_2$D$^+$, N$_2$D$^+$, and DCO$^+$.

\subsection{Chance superposition cores and other line-of-sight gas components seen towards Orion~B9}

While the LSR velocities of the detected \textit{o}-H$_2$D$^+$ lines are about 9~km~s$^{-1}$, the \textit{o}-H$_2$D$^+$ emission 
from the prestellar core SMM~7 appears at $v_{\rm LSR}\simeq3.6$~km~s$^{-1}$. This is also the case for the detected DCO$^+(5-4)$ lines 
(see Appendix~A). Indeed, Miettinen et al. (2010, 2012) found that the NH$_3(1,\,1)$, NH$_3(2,\,2)$, C$^{17}$O$(2-1)$, DCO$^+(4-3)$, and N$_2$H$^+(3-2)$ line emissions towards SMM~7 occur at 3.6-4~km~s$^{-1}$. These results suggest that SMM~7 might be a chance projection seen towards the Orion~B9 filament.  

Moreover, the protostellar source IRAS~05413-0104 that lies $13\farcm2$ (1.6~pc) to the northeast of SMM~7 (see Fig.~1 in \cite{miettinen2009}) was found to have NH$_3$ radial velocities that are significantly lower than the bulk of the Orion~B9 cores, namely $\sim1.5$~km~s$^{-1}$ (\cite{miettinen2010}). Also, the protostellar core SMM~4, which was not chosen for the present study, was found to exhibit line emission at 1.5-1.7~km~s$^{-1}$ by Miettinen et al. (2012), although the NH$_3$ lines detected towards SMM~4 showed two components, one at the systemic velocity of Orion~B9 ($\sim9$~km~s$^{-1}$) and the other at about 1.6~km~s$^{-1}$. This could be caused by the large, $40\arcsec$ beam of the NH$_3$ observations that also captured the 9~km~s$^{-1}$ gas component (see Fig.~\ref{figure:map}). Therefore, in addition to SMM~7, the protostellar cores SMM~4 and IRAS~05413 may also be physically unrelated to the Orion~B9 star-forming region. 

One of the present target sources, IRAS~05405, exhibits additional velocity components in the C$^{17}$O$(2-1)$ and 
N$_2$D$^+(3-2)$ spectra (\cite{miettinenetal2012}). As was pointed out by Miettinen et al. (2010), the dense gas and dust associated with the additional line-of-sight velocity components can also affect the physical parameters of the cores derived from dust continuum observations. For example, the dust-based H$_2$ column density can be overestimated if the dust continuum surface brightness is 
contaminated by a physically unrelated component. This would then lead to underestimated fractional abundances of the detected molecules. For instance, the lower velocity C$^{17}$O$(2-1)$ lines detected towards IRAS~05405 (at 1.3~km~s$^{-1}$ and 3.0~km~s$^{-1}$) are $\sim60-70\%$ of the intensity of the systemic velocity (9.2~km~s$^{-1}$) line (\cite{miettinenetal2012}; Table~5 therein). However, the beam size of the aforementioned C$^{17}$O$(2-1)$ observations was 1.65 times larger than that of the present \textit{o}-H$_2$D$^+$ observations, and therefore the contribution from the additional line-of-sight velocity components to the present \textit{o}-H$_2$D$^+$ abundance upper limit is unclear. Moreover, the H$_2$ column density estimates suffer from other uncertain parameters, especially from the uncertainty in the submillimetre dust opacity that can be a factor of approximately two (e.g. \cite{shirley2011}). In conclusion, the additional velocity components seen towards Orion~B9 are relevant for only one of the present targets (IRAS~05405), but their relative contribution to the submillimetre dust continuum emission is difficult to reliably quantify. High-resolution molecular line imaging would be useful to resolve this issue.

\section{Summary and conclusions}

We used the APEX telescope to observe the 372~GHz \textit{o}-H$_2$D$^+(J_{K_a,\,K_c}=1_{1,\,0}-1_{1,\,1})$ line towards three prestellar cores and three protostellar cores in the Orion~B9 filament. The \textit{o}-H$_2$D$^+$ data were analysed in conjunction with our previous APEX spectral line and dust continuum data for the target sources. Our main results are summarised as follows:

\begin{enumerate}
\item The \textit{o}-H$_2$D$^+(J_{K_a,\,K_c}=1_{1,\,0}-1_{1,\,1})$ line was detected in all three prestellar cores with the abundances in the range $x(o-{\rm H_2D^+})\sim(12-30)\times10^{-11}$. Only one of the protostellar cores, the Class~0 object SMM~3, was detected in \textit{o}-H$_2$D$^+$ emission with an abundance of $6\times10^{-11}$. 
\item Besides the \textit{o}-H$_2$D$^+$ line detections, the N$_2$H$^+(4-3)$ line was detected towards all the sources except one prestellar core (SMM~7), and DCO$^+(5-4)$ was detected in all the target sources. 
\item No significant correlations were found between the level of deuterium fractionation (quantified as the $[{\rm N_2D^+}]/[{\rm N_2H^+}]$ ratio), the factor of CO depletion, the gas kinetic temperature derived from ammonia, or the \textit{o}-H$_2$D$^+$ abundance. However, our results are in fairly good agreement with the $x(o-{\rm H_2D^+})-T_{\rm kin}$ relationship derived by Caselli et al. (2008) for low-mass dense cores.
\item The derived \textit{o}-H$_2$D$^+$ abundances in Orion~B9 are more similar to those observed in other low-mass star-forming regions than to those derived for high-mass star-forming clumps where the values are typically a few times $10^{-11}$ or less.
\item We found that the \textit{o}-H$_2$D$^+$ abundance appears to decrease as a function of temporal core evolution. We also addressed the evolutionary scenario proposed by Giannetti et al. (2019), namely that the 
[\textit{o}-${\rm H_2D^+}]/[{\rm N_2D^+}]$ abundance ratio decreases as the core evolves towards more advanced stages. However, our results were not conclusive, which might (partly) be caused by the spatial offset between our \textit{o}-H$_2$D$^+$ and N$_2$D$^+$ observations 
($\sim10\arcsec$ on average). Indeed, the N$_2$D$^+$ abundance was not found to increase as the core evolves, but it rather showed a fluctuating behaviour. On the other hand, the [\textit{o}-${\rm H_2D^+}]/[{\rm DCO^+}]$ ratio, where the DCO$^+$ data were observed simultaneously with \textit{o}-H$_2$D$^+$, was also not found to exhibit any clear evolutionary trend.
\item The prestellar core SMM~7 exhibits \textit{o}-H$_2$D$^+$ line emission at a radial velocity that is almost 
$\sim6$~km~s$^{-1}$ lower than the systemic velocity of Orion~B9. Therefore, the source might be a chance projection seen towards the 
Orion~B9 filament region.
\end{enumerate}

Orion~B9 is part of the dynamic Orion~B environment, and could itself be a 
region of triggered core and star formation that lies about $\sim4$~pc from the \ion{H}{II} region 
NGC~2024. However, large-scale molecular line mapping (covering at least the whole Orion~B9 filament) is required to better understand the line-of-sight velocity distribution of Orion B9, and to quantitatively test the hypothesis that the region 
is indeed affected by feedback processes. If confirmed, the Orion~B9 filament provides an interesting target system
to investigate the deuterium-based chemistry and the possible environmental effects 
in an important class of star-forming regions. Observations of the \textit{para} form of H$_2$D$^+$ would be useful 
to constrain the ages of the cores (\cite{brunken2014}; \cite{harju2017}), and thus the age of the whole parent filament. 
Also, to better understand the behaviour of the [\textit{o}-${\rm H_2D^+}]/[{\rm N_2D^+}]$ evolutionary indicator proposed by 
Giannetti et al. (2019), observations of D$_2$H$^+$ would be useful to investigate the importance of 
the deuteration sequence ${\rm H_3^+}\rightarrow {\rm H_2D^+}\rightarrow {\rm D_2H^+}$ in dense cores destined to become new stars.

\begin{acknowledgements}

I thank the referee for the critical comments and suggestions. 
I am grateful to the staff at the APEX telescope for performing the service mode LAsMA 
observations presented in this paper. I would also like to thank S\'ebastien Bardeau and S\'ebastien Maret 
for the help with the GILDAS software package. This research has made use of NASA's Astrophysics Data System Bibliographic Services. 
This research made use of {\tt Astropy}\footnote{\url{http://www.astropy.org}}, a community-developed core Python package for Astronomy (\cite{astropy2013}, \cite{astropy2018}).

\end{acknowledgements}

\appendix

\section{Other line detections}

The frequency bands covered by our APEX/LAsMA observations allowed 
us to detect additional spectral lines in the target sources. 
To identify these lines, we used the {\tt Weeds} interface, which is an
extension of {\tt CLASS} (\cite{maret2011}), to access the CDMS  
spectroscopic database.

\begin{figure*}[!htb]
\begin{center}
\includegraphics[width=0.33\textwidth]{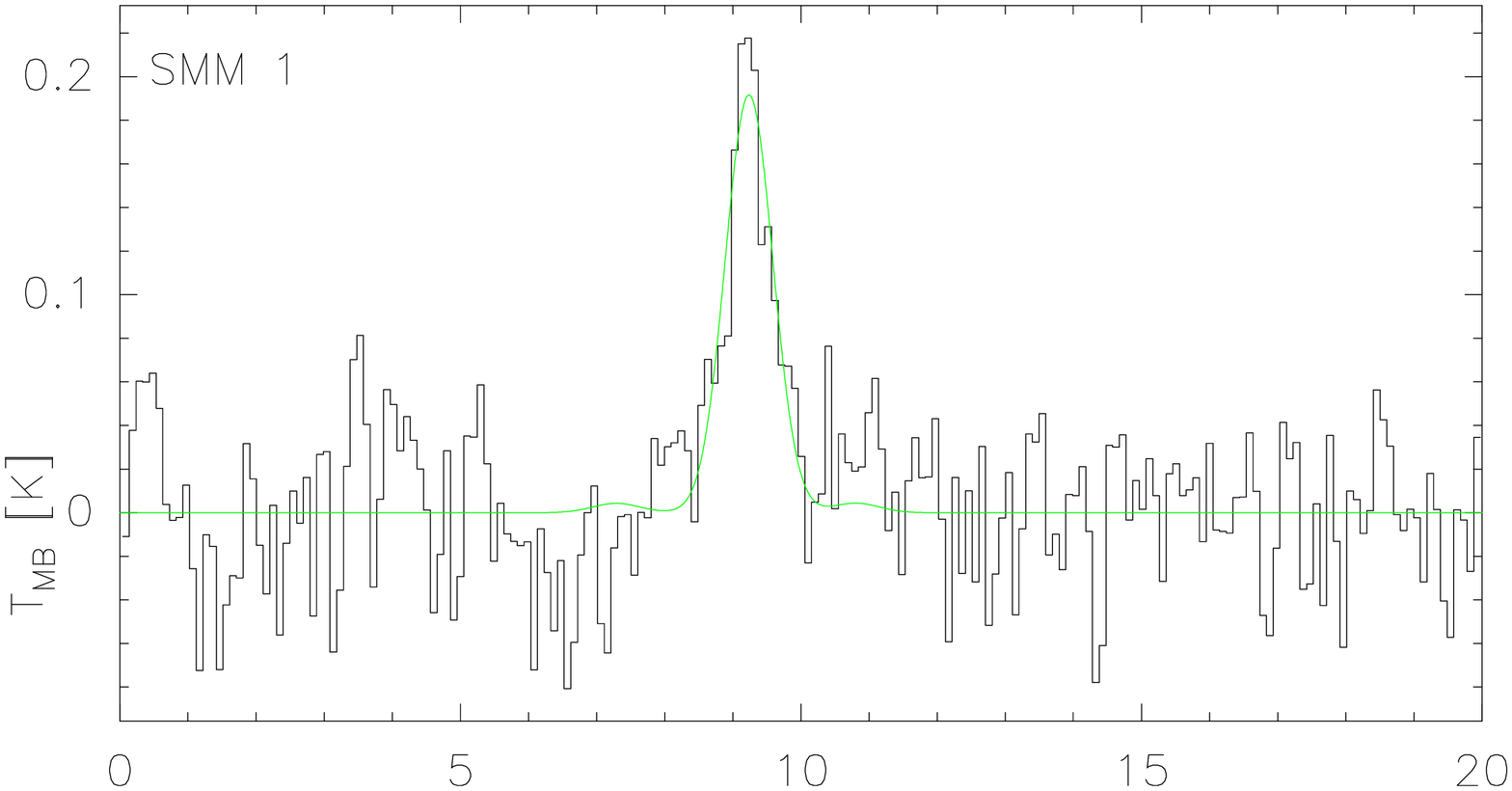}
\includegraphics[width=0.33\textwidth]{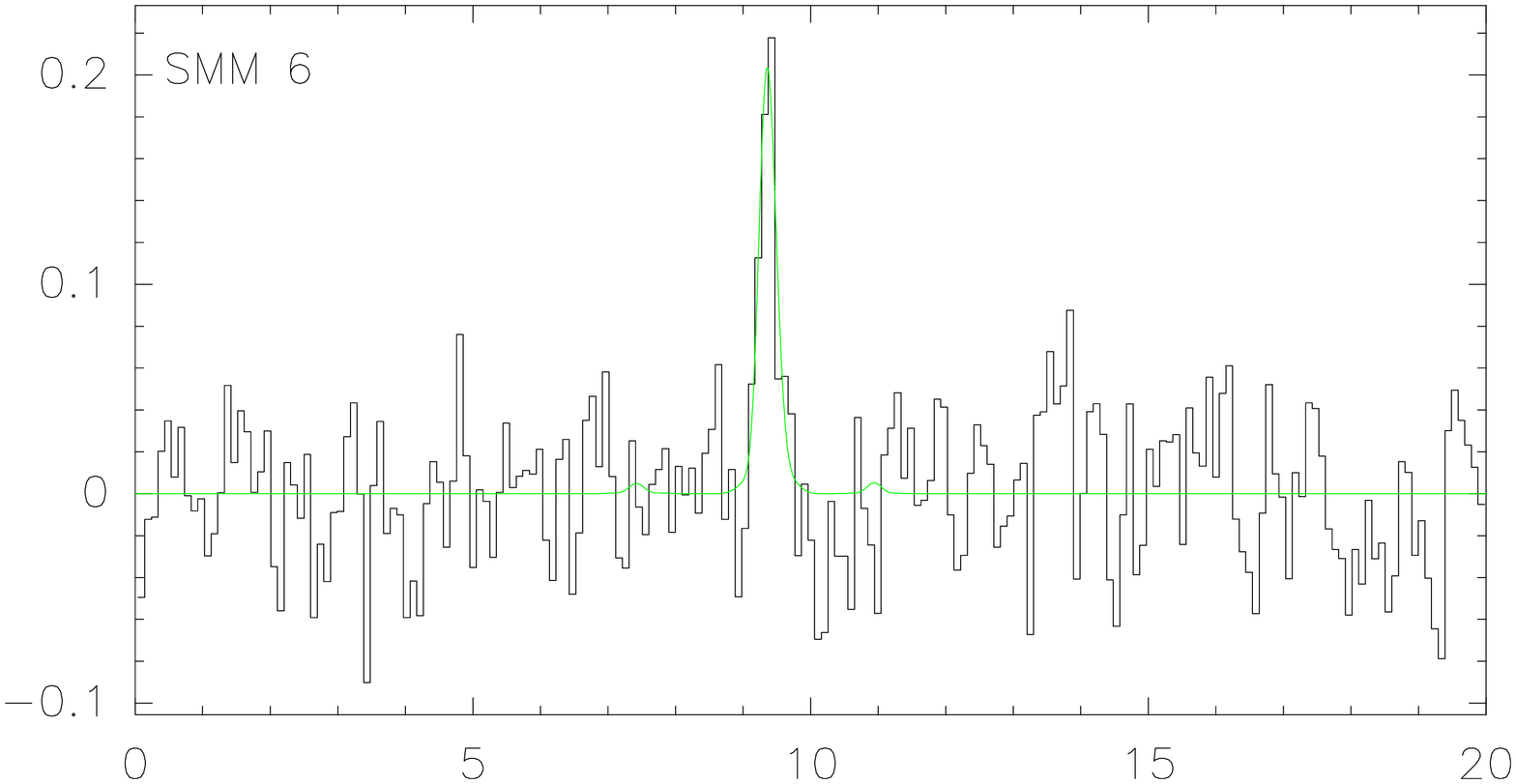}
\includegraphics[width=0.33\textwidth]{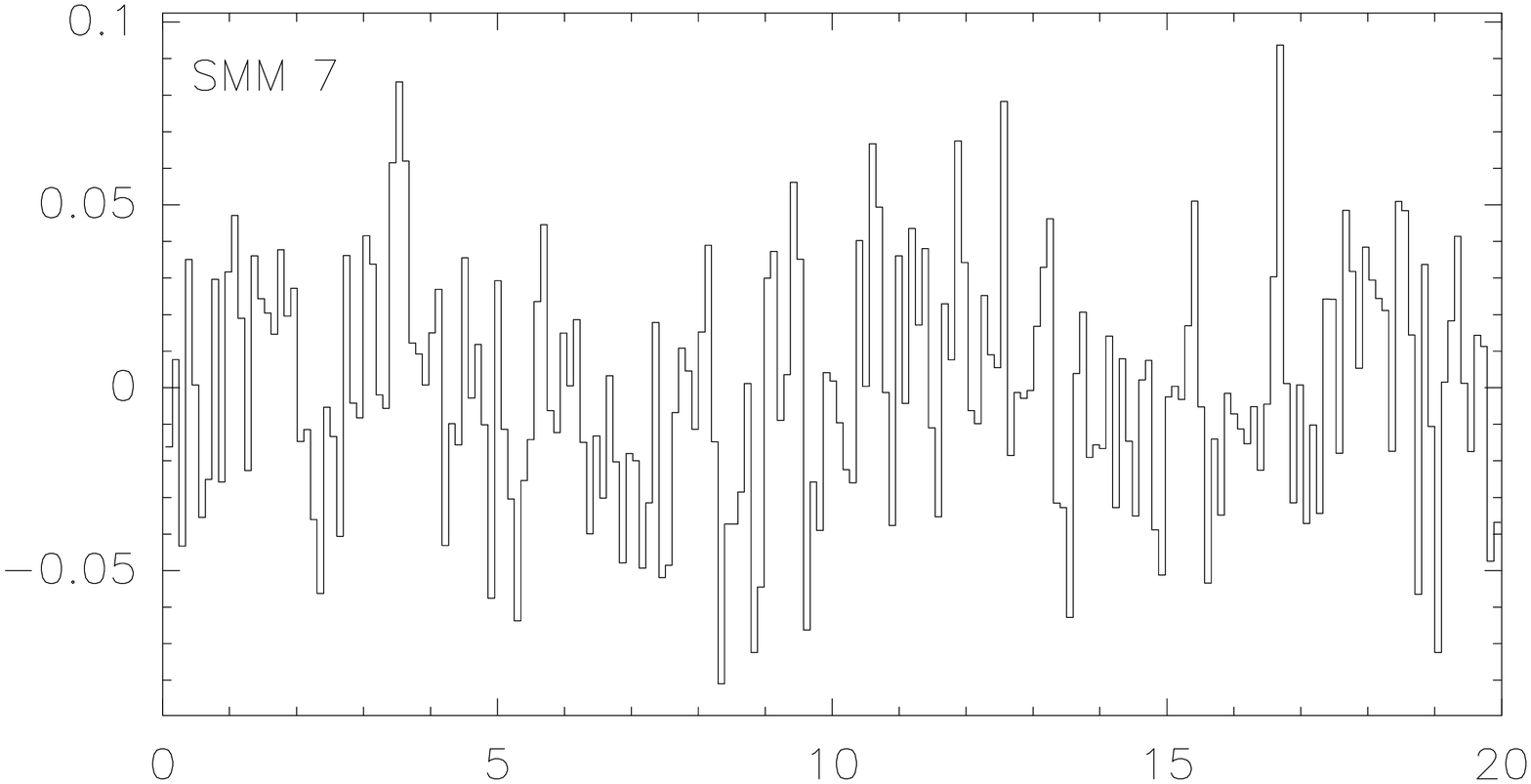}
\includegraphics[width=0.33\textwidth]{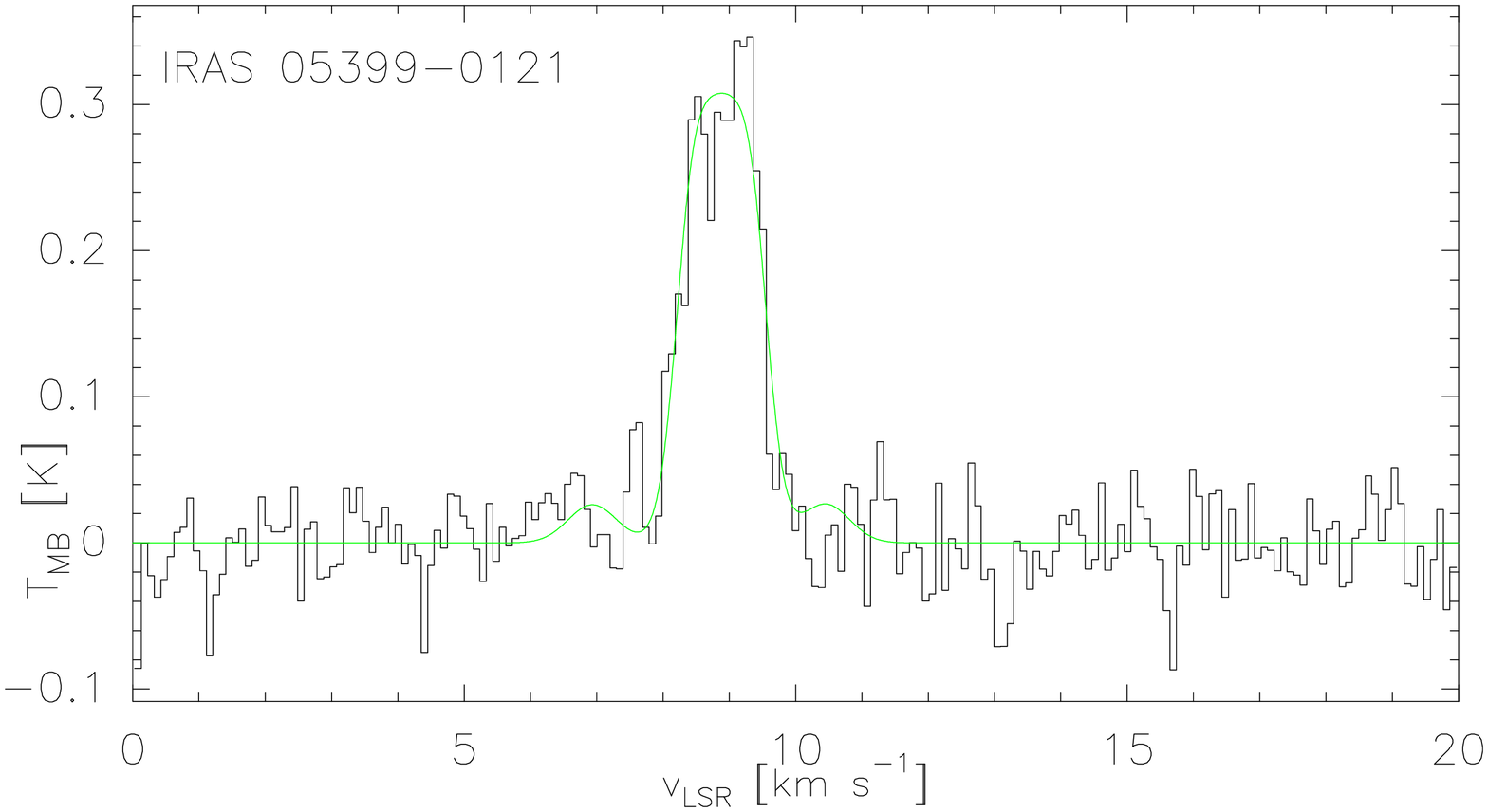}
\includegraphics[width=0.33\textwidth]{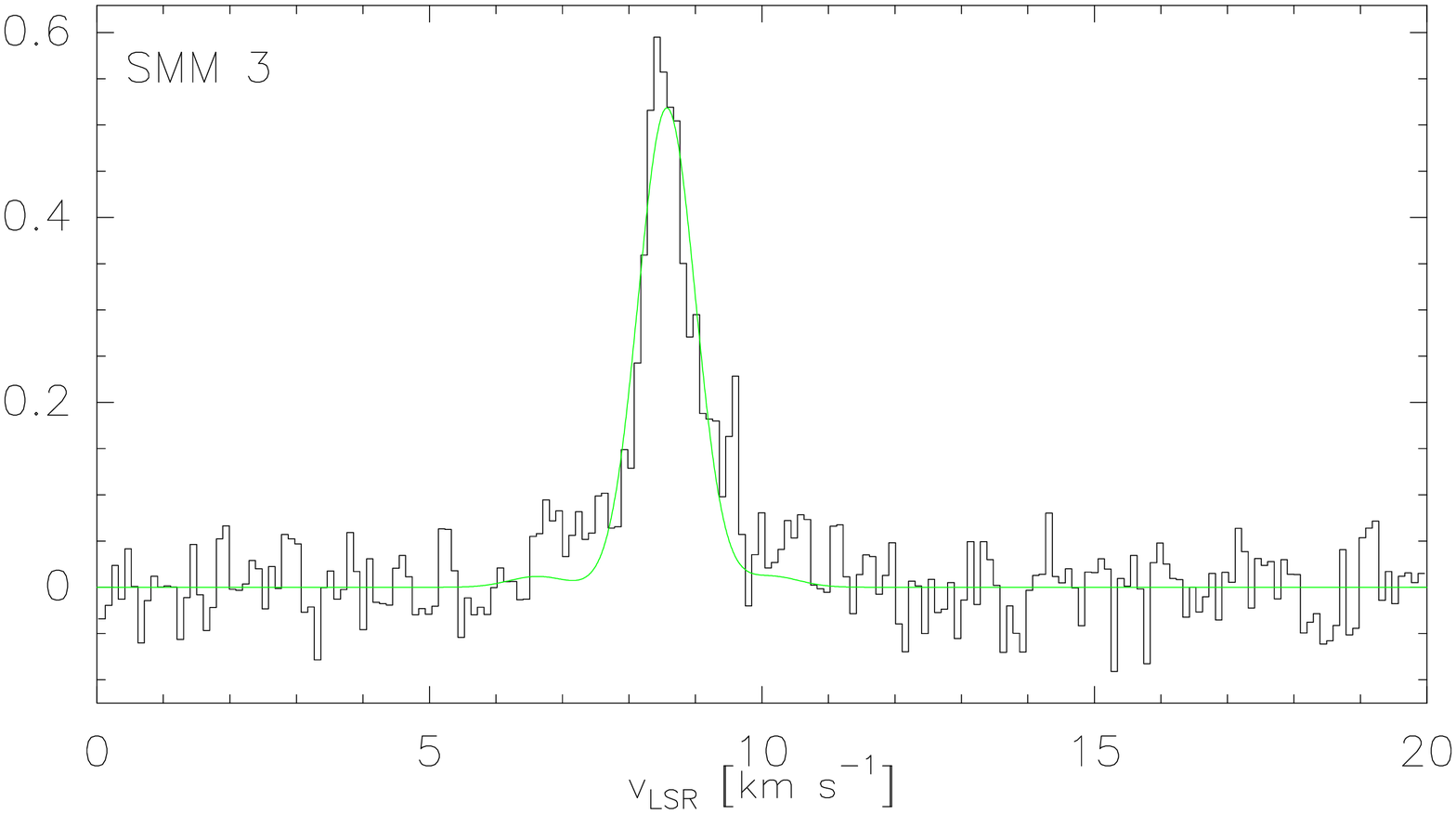}
\includegraphics[width=0.33\textwidth]{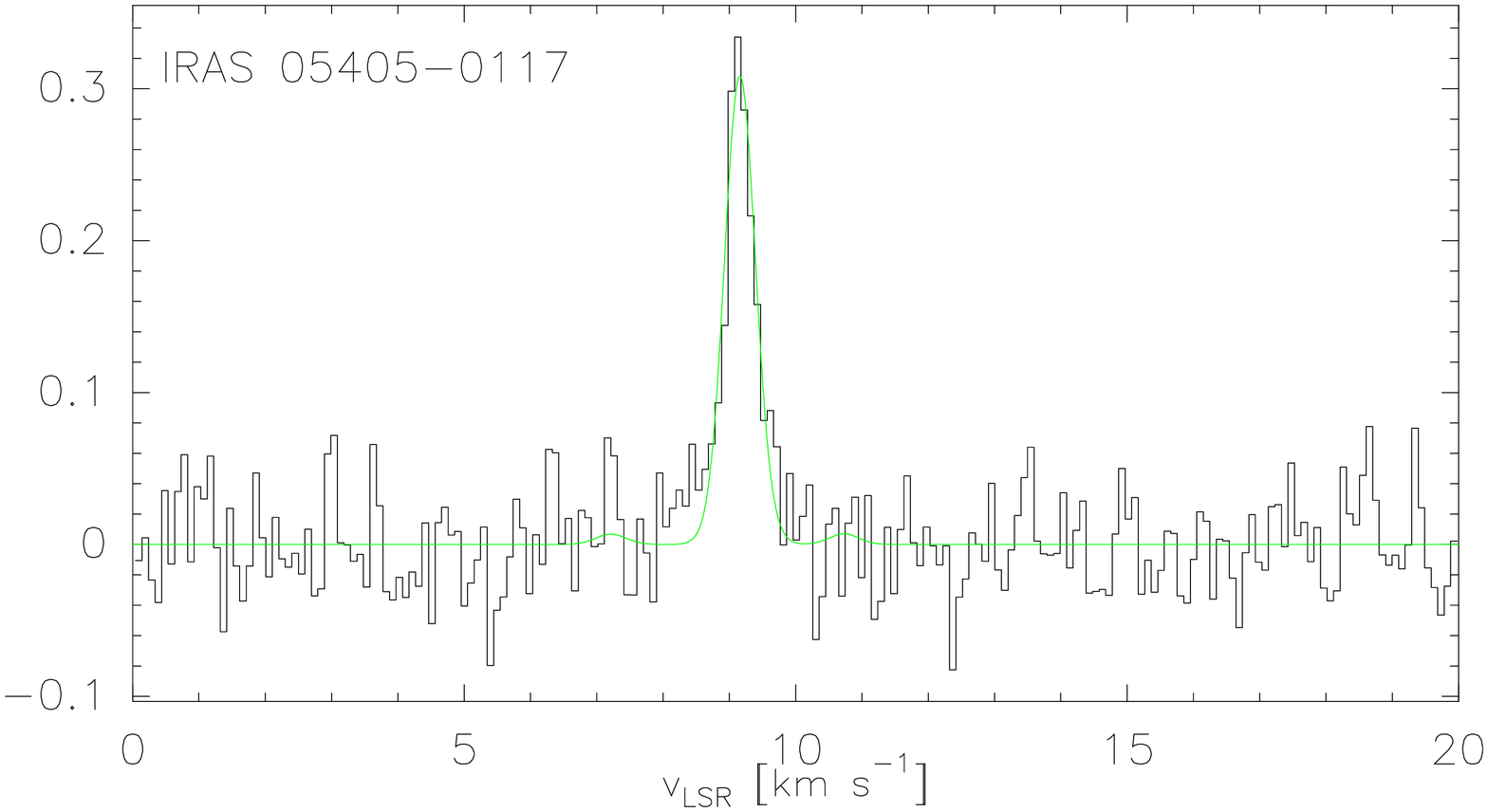}
\caption{APEX N$_2$H$^+(4-3)$ spectra towards the prestellar cores (\textit{top row}) and protostellar cores (\textit{bottom row}) in our sample. Hyperfine structure fits to the lines are overlaid in green. While the velocity range shown in each panel is the same, the intensity range is different to better show the line profiles.}
\label{figure:otherspectra1}
\end{center}
\end{figure*}

\begin{figure*}[!htb]
\begin{center}
\includegraphics[width=0.33\textwidth]{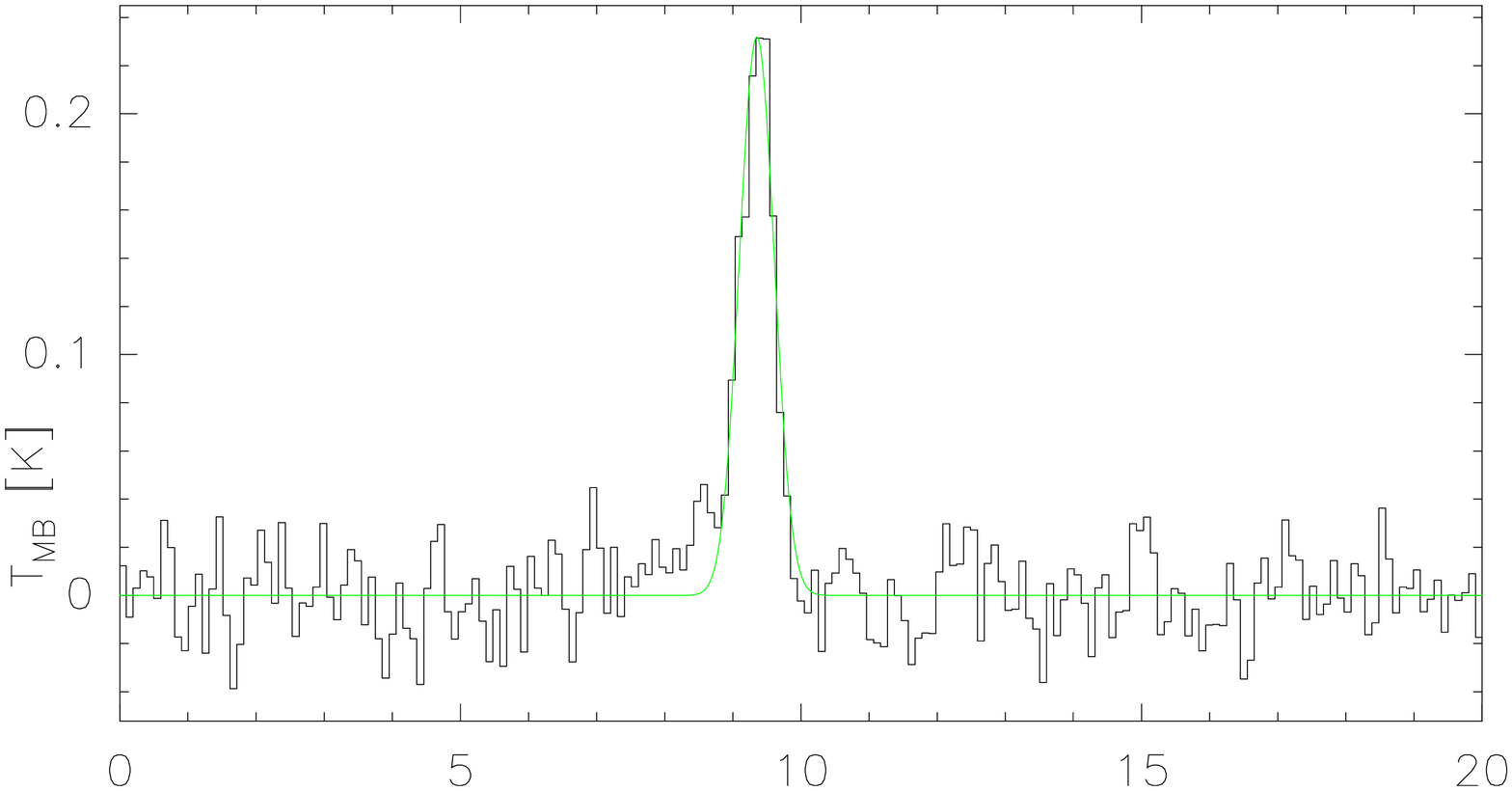}
\includegraphics[width=0.33\textwidth]{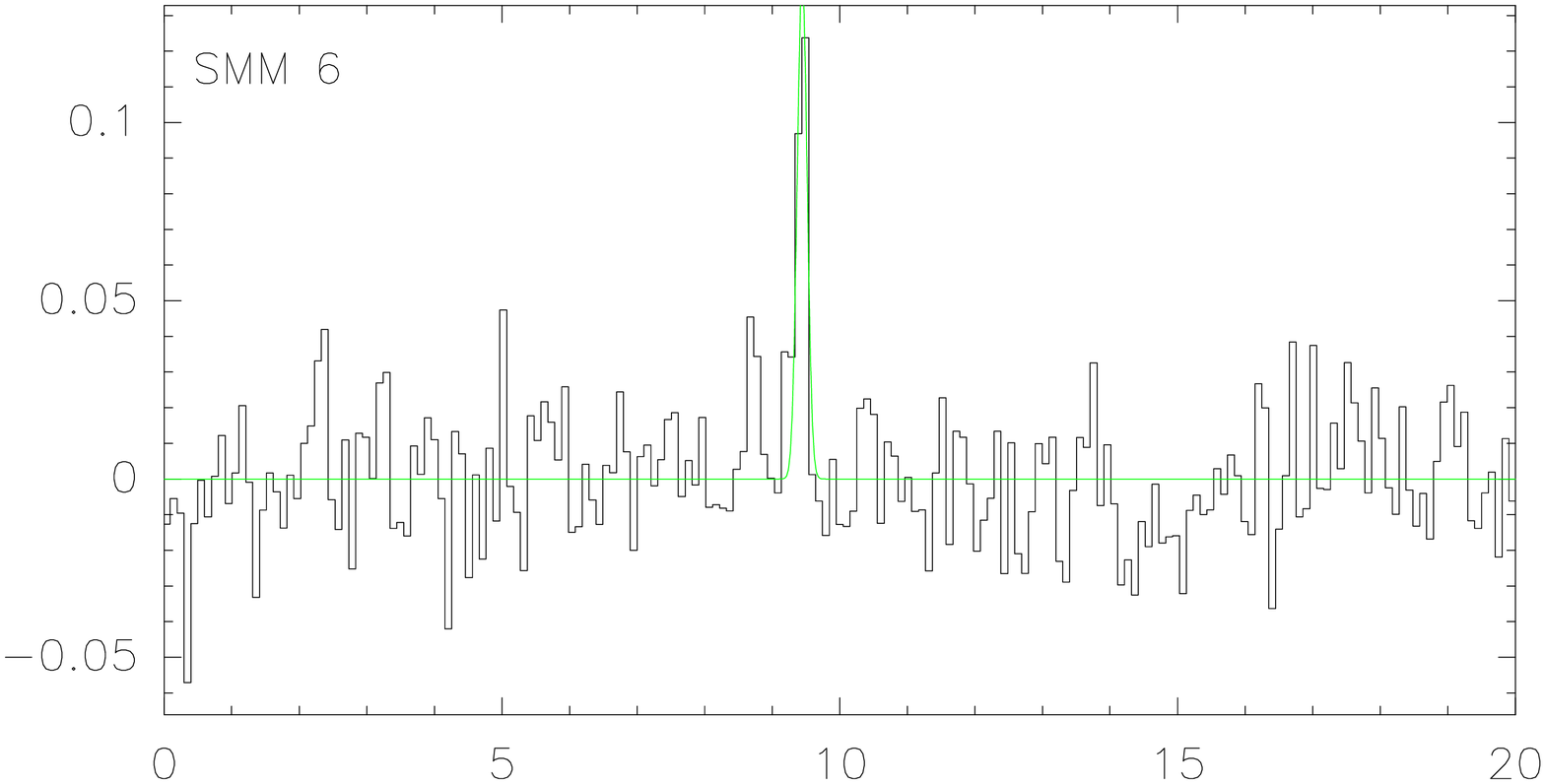}
\includegraphics[width=0.33\textwidth]{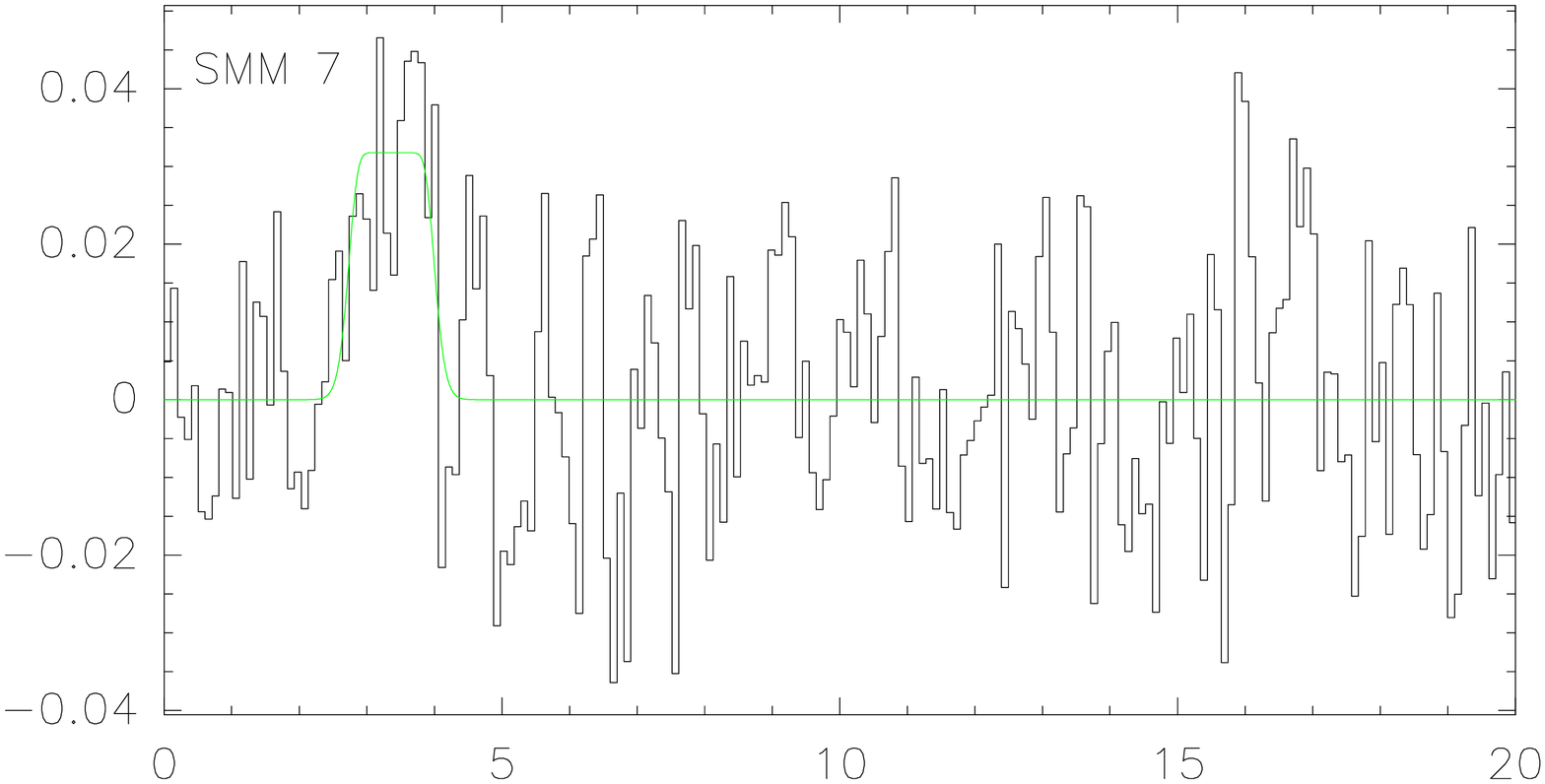}
\includegraphics[width=0.33\textwidth]{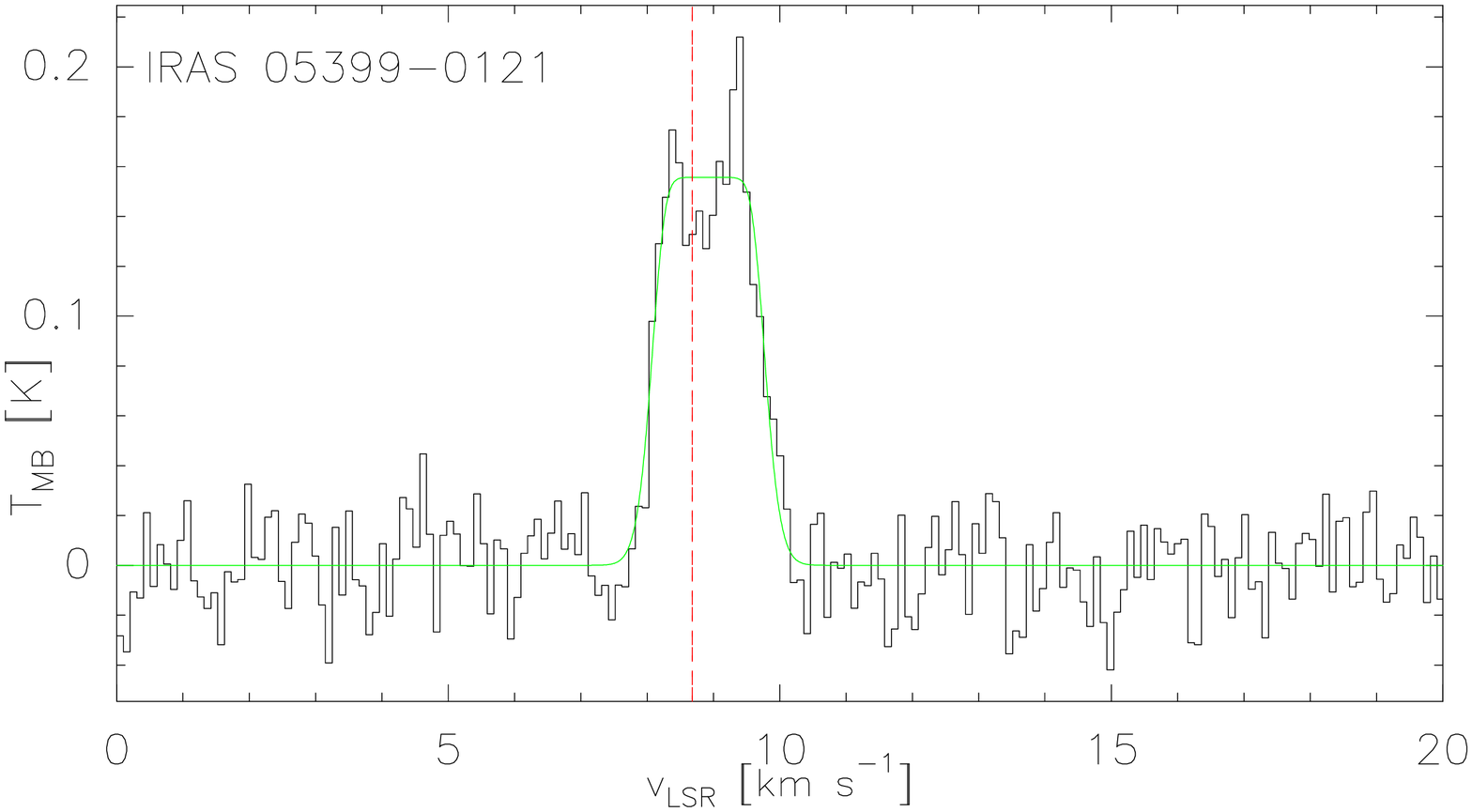}
\includegraphics[width=0.33\textwidth]{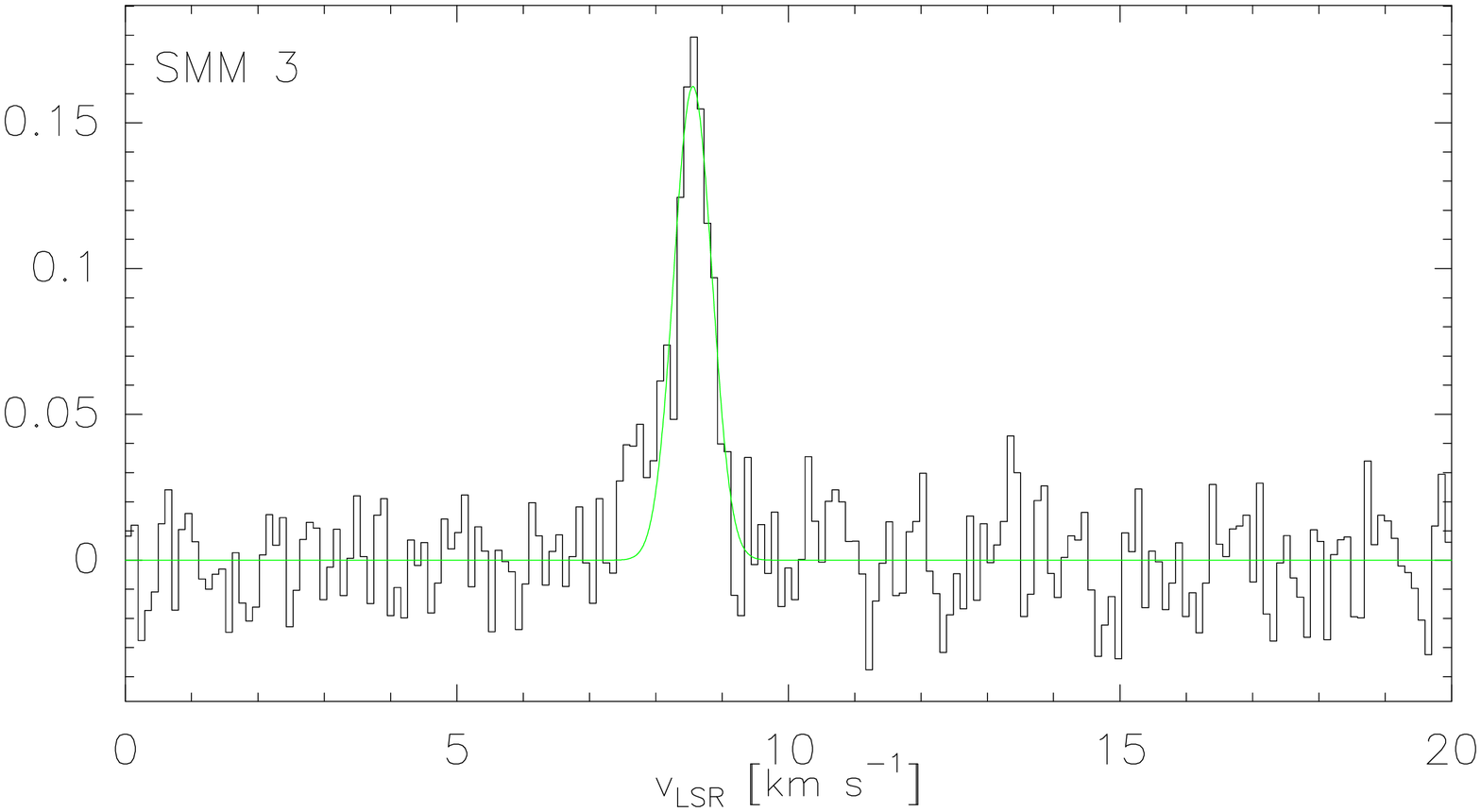}
\includegraphics[width=0.33\textwidth]{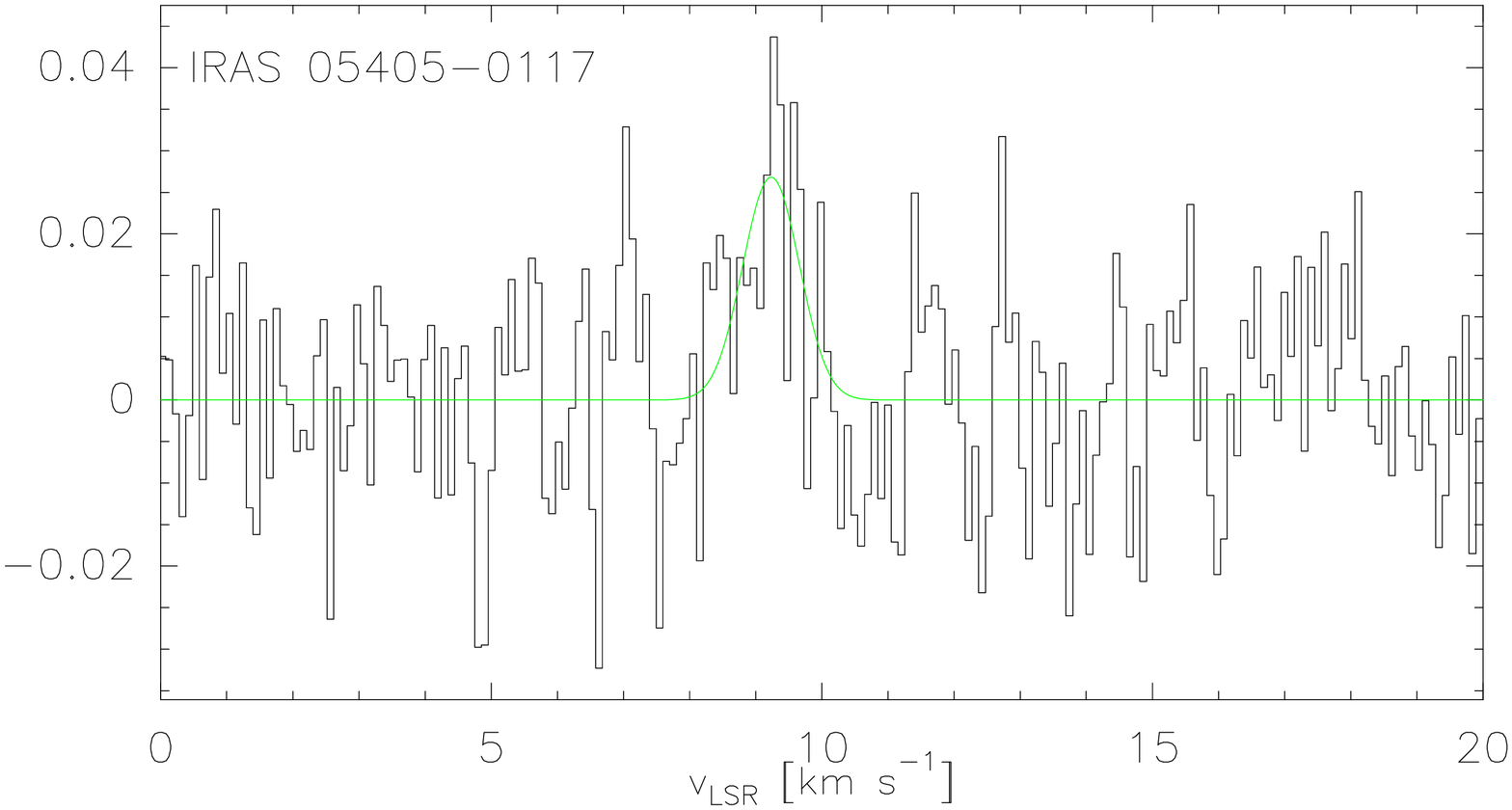}
\caption{APEX DCO$^+(5-4)$ spectra towards the prestellar cores (\textit{top row}) and protostellar cores (\textit{bottom row}) in our sample. Hyperfine structure fits to the lines are overlaid in green. While the velocity range shown in each panel is the same, the intensity range is different to better show the line profiles. A double-peaked profile with a stronger red peak is detected towards IRAS~05399. The red, vertical dashed line in the IRAS~05399 panel shows the systemic velocity derived from C$^{17}$O$(2-1)$ by Miettinen et al. (2012).}
\label{figure:otherspectra2}
\end{center}
\end{figure*}

Two additional spectral lines could be identified, the $J=4-3$ transition of N$_2$H$^+$ 
and $J=5-4$ transition of DCO$^+$, where the former was detected in all sources except SMM~7, 
while the latter one was detected in all six sources. The spectra are shown in 
Figs.~\ref{figure:otherspectra1} and \ref{figure:otherspectra2}. We note that in the case of IRAS~05399, 
the LSB data observed on 15-16 December 2019 were omitted from the summed spectrum owing to their disruptive effect 
on the line profile (i.e. the data strongly affected the appearance of the double-peaked DCO$^+(5-4)$ profile). The poorer 
quality of the December data compared to those taken in August was potentially caused by the up to $\sim70\%$ higher amount of PWV (see Table~\ref{table:observations}). After the removal of the aforementioned December data, the effective on-source time for the DCO$^+(5-4)$ observations towards IRAS~05399 was 22.4~min.

The $J=4-3$ transition of N$_2$H$^+$ is split into 38 hyperfine components. We fit this hyperfine structure 
using the rest frequencies from Pagani et al. (2009b, Table~5 therein) and the {\tt CLASS} program.
The adopted central frequency of N$_2$H$^+(4-3)$, 372\,672.526~MHz, is 
that of the strongest hyperfine component, $J_{F_1F} = 4_{56} \rightarrow 3_{45}$, 
which has a relative intensity of $R_i = 13/49$. Owing to the 
heavy blending of the hyperfine components, they could not be used to reliably 
determine the line optical thickness. 

The $J=5-4$ transition of DCO$^+$ is split into six hyperfine components, 
and hence the detected lines were fit using the hyperfine structure method of {\tt CLASS}. 
The hyperfine component frequencies were taken from the Jet Propulsion Laboratory (JPL) spectroscopic database\footnote{\url{http://spec.jpl.nasa.gov/}} (\cite{pickett1998}). The central frequency was taken to be 360\,169.7771~MHz, which corresponds 
to that of the strongest hyperfine component $F=6-5$ with a relative intensity of $R_i = 13/33$. 
The derived line parameters are tabulated in Table~\ref{table:otherparameters}. 

In the last two columns in Table~\ref{table:otherparameters}, we also list the N$_2$H$^+$ and DCO$^+$ column densities 
and fractional abundances derived from the aforementioned transitions. The analysis was similar to that described in Sect.~3.2 
(see Eq.~(\ref{eqn:CD}) therein). The $T_{\rm ex}[{\rm NH_3(1,\, 1)}]$ values were used for both N$_2$H$^+$ and DCO$^+$. An upper limit to the N$_2$H$^+$ column density for SMM~7 was estimated using the N$_2$H$^+(3-2)$ line width ($0.67\pm0.27$~km~s$^{-1}$; \cite{miettinenetal2012}). We note that the N$_2$H$^+$ column densities derived from the $J=3-2$ and $J=4-3$ transitions towards the same position in SMM~6 are in fairly good agreement within the uncertainties (their ratio is $0.7\pm0.2$; see Table~\ref{table:results}). The present DCO$^+$ column density derived for SMM~3 is $2.8\pm1.2$ times higher than that derived from the $J=3-2$ line towards the core centre by Miettinen (2016; Table~4 therein).

\begin{table*}
\caption{Additional lines detected towards the target sources.}
{\small
\begin{minipage}{2\columnwidth}
\centering
\renewcommand{\footnoterule}{}
\label{table:otherparameters}
\begin{tabular}{c c c c c c c c c}
\hline\hline 
Source & Transition & $v_{\rm LSR}$ & $\Delta v_{\rm LSR}$ & $T_{\rm MB}$ & $\int T_{\rm MB} {\rm d}v$ & $\tau$ & $N$ & $x$\\
       &            & [km s$^{-1}$] & [km s$^{-1}$] & [K] & [K km s$^{-1}$] & & [$10^{13}$ cm$^{-2}$] & [$10^{-10}$]\\
\hline
IRAS 05399-0121 & N$_2$H$^+(J=4-3)$ & $8.82\pm0.02$ & $0.81\pm0.05$ & $0.33\pm0.04$ & $0.44\pm0.05$ & $2.66\pm0.34$ & $16.3\pm2.3$ & $26.3\pm4.7$\\ [1ex] 
				& DCO$^+(J=5-4)$ & $8.93\pm0.02$ & $0.85\pm0.06$ & $0.18\pm0.02$ & $0.29\pm0.03$ & $0.75\pm0.11$ & $15.6\pm2.5$ & $25.0\pm4.9$ \\[1ex] 
				& \ldots & $8.37\pm0.03$\tablefootmark{a} & $0.60\pm0.06$ & $0.16\pm0.02$ & $0.10\pm0.02$ & \ldots & \ldots & \ldots\\ [1ex] 
				& \ldots & $9.29\pm0.04$\tablefootmark{b} & $0.90\pm0.08$ & $0.18\pm0.03$ & $0.18\pm0.02$ & \ldots & \ldots & \ldots\\ [1ex] 
SMM 1 & N$_2$H$^+(J=4-3)$ & $9.17\pm0.04$ & $0.78\pm0.10$ & $0.19\pm0.04$ & $0.17\pm0.02$ & $0.48\pm0.04$ & $1.1\pm0.2$ & $2.6\pm0.5$\\ [1ex] 
	  & DCO$^+(J=5-4)$ & $9.36\pm0.01$ & $0.59\pm0.03$ & $0.23\pm0.03$ & $0.15\pm0.01$ & $0.37\pm0.04$ & $1.5\pm0.2$ & $3.4\pm0.6$\\[1ex] 
SMM 3 & N$_2$H$^+(J=4-3)$ & $8.51\pm0.02$ & $0.97\pm0.06$ & $0.51\pm0.06$ & $0.56\pm0.06$ & $2.76\pm0.28$ & $13.0\pm1.6$ & $11.9\pm1.9$\\ [1ex] 
      & DCO$^+(J=5-4)$ & $8.56\pm0.02$ & $0.66\pm0.05$ & $0.16\pm0.02$ & $0.12\pm0.01$ & $0.42\pm0.05$ & $3.7\pm0.5$ & $3.4\pm0.6$\\[1ex] 
IRAS 05405-0117 & N$_2$H$^+(J=4-3)$ & $9.10\pm0.02$ & $0.51\pm0.05$ & $0.30\pm0.04$ & $0.18\pm0.03$ & $0.51\pm0.04$ & $0.5\pm0.1$ & $3.0\pm0.7$ \\ [1ex] 
      & DCO$^+(J=5-4)$ & $9.24\pm0.12$ & $0.99\pm0.24$ & $0.03\pm0.01$ & $0.03\pm0.01$ & $0.03\pm0.01$ & $0.1\pm0.03$ & $0.7\pm0.2$\\[1ex] 
SMM 6 & N$_2$H$^+(J=4-3)$ & $9.31\pm0.02$ & $0.25\pm0.05$ & $0.21\pm0.04$ & $0.07\pm0.01$ & $0.81\pm0.06$ & $0.9\pm0.2$ & $3.2\pm0.9$\\ [1ex]      
      & DCO$^+(J=5-4)$ & $9.45\pm0.01$ & $0.19\pm0.04$ & $0.14\pm0.02$ & $0.03\pm0.01$ & $0.33\pm0.03$ & $0.7\pm0.2$ & $2.6\pm0.7$\\[1ex] 
SMM 7 & N$_2$H$^+(J=4-3)$ & \ldots & \ldots & $<0.09$ & $<0.09$\tablefootmark{c} & $<2.36$ & $<58$\tablefootmark{c} & $<142$\tablefootmark{c} \\ [1ex] 
      & DCO$^+(J=5-4)$ & $3.37\pm0.08$ & $0.60\pm0.08$ & $0.04\pm0.02$ & $0.04\pm0.01$ & $0.49\pm0.05$ & $41.5\pm7.0$ & $99.9\pm22.9$\\[1ex] 
\hline 
\end{tabular} 
\tablefoot{\tablefoottext{a}{The parameters on this row were derived from a Gaussian fit to the blueshifted peak.}\tablefoottext{b}{The parameters on this row were derived from a Gaussian fit to the redshifted peak.}\tablefoottext{c}{These upper limits were calculated by using the quoted $3\sigma$ intensity upper limit and the FWHM of the N$_2$H$^+(3-2)$ line ($0.67\pm0.27$~km~s$^{-1}$) detected by Miettinen et al. (2012).}}
\end{minipage}}
\end{table*}

\end{document}